\pdfoutput=1
\documentclass[11pt]{article}

\usepackage[preprint]{acl}
\usepackage{times}
\usepackage{latexsym}
\usepackage{array}
\usepackage[T1]{fontenc}
\usepackage[utf8]{inputenc}

\usepackage{microtype}

\usepackage{inconsolata}

\usepackage{times}
\usepackage{latexsym}
\usepackage{graphicx}
\usepackage{hyperref, hyperxmp}
\usepackage{svg}
\usepackage{amsmath}
\usepackage{tikz}
\usepackage{pifont}
\usepackage{wasysym}
\usepackage{mathtools}
\usepackage{lipsum}% http://ctan.org/pkg/lipsum
\usepackage{graphicx}% http://ctan.org/pkg/graphicx
\usepackage{booktabs}
\usepackage{caption, subcaption}
\usepackage{url}
\usepackage{amsmath, amsfonts, amsthm, bm}
\usepackage{soul}
\usepackage{tabularx, multirow}
\usepackage{dsfont}
\usepackage[ruled,vlined,linesnumbered]{algorithm2e}
\usepackage{booktabs}
\usepackage{xcolor, colortbl}
\usepackage{xspace}
\usepackage{resizegather}
\usepackage{makecell}
\usepackage[normalem]{ulem}
\usepackage{CJKutf8}
\usepackage{enumitem}
\usepackage{arydshln}
\usepackage{enumitem}
\usepackage{balance}

\title{Why These Documents? Explainable Generative Retrieval with Hierarchical Category Paths}

\author{
Sangam Lee$^1$, Ryang Heo$^1$, SeongKu Kang$^2$, Susik Yoon$^2$, \\\textbf{Jinyoung Yeo}$^1$, \textbf{Dongha Lee}$^1$\thanks{Corresponding author}\\
  $^1$Yonsei University \qquad  $^2$Korea University \\
  \texttt{\{salee,ryang1119,jinyeo,donalee\}@yonsei.ac.kr}\\
  \texttt{\{seongkukang,susik\}@korea.ac.kr}\\}

\definecolor{DarkGreen}{RGB}{30,130,30}

\definecolor{DarkYellow}{RGB}{255,204,0}

\newcommand{\hype}{\textsc{HyPE}\xspace}
\newcommand{\pars}{path-aware ranking\xspace}

\newcommand{\gr}{generative retrieval\xspace}

\newcommand{\llm}{LLM\xspace}

\newcommand{\textbasedocid}{lexical docid\xspace}

\newcommand{\nq}{NQ320K\xspace}
\newcommand{\msmarco}{MS MARCO\xspace}

\begin{document}
\maketitle
\begin{abstract}
Generative retrieval directly decodes a document identifier (i.e., docid) in response to a query, making it impossible to provide users with explanations as an answer for \textit{``why is this document retrieved?''}. 
To address this limitation, we propose \underline{\textbf{H}}ierarchical Categor\underline{\textbf{y}} \underline{\textbf{P}}ath-\underline{\textbf{E}}nhanced Generative Retrieval (\textbf{\hype}), which enhances explainability by first generating hierarchical category paths step-by-step then decoding docids.
By leveraging hierarchical category paths which progress from broader to more specific semantic categories, \hype can provide detailed explanations for its retrieval decision. 
For training, \hype constructs category paths with external high-quality semantic hierarchy, leverages \llm to select appropriate candidate paths for each document, and optimizes the generative retrieval model with path-augmented dataset. 
During inference, \hype utilizes \pars strategy to aggregate diverse topic information, allowing the most relevant documents to be prioritized in the final ranked list of docids.
Our extensive experiments demonstrate that \hype not only offers a high level of explainability but also improves the retrieval performance. 
We provide the code and a live demo of \hype at \url{https://augustinlib.github.io/HyPE/}.
\end{abstract}

\section{Introduction}
\label{sec:intro}
Information retrieval systems are essential for helping users find proper information within a vast amount of online information. 
A fundamental task of these systems is document retrieval, which focuses on searching for and ranking documents that are relevant to a given query from a large corpus. 
\begin{figure}[!tb]
    \centering
        \includegraphics[width=\columnwidth]{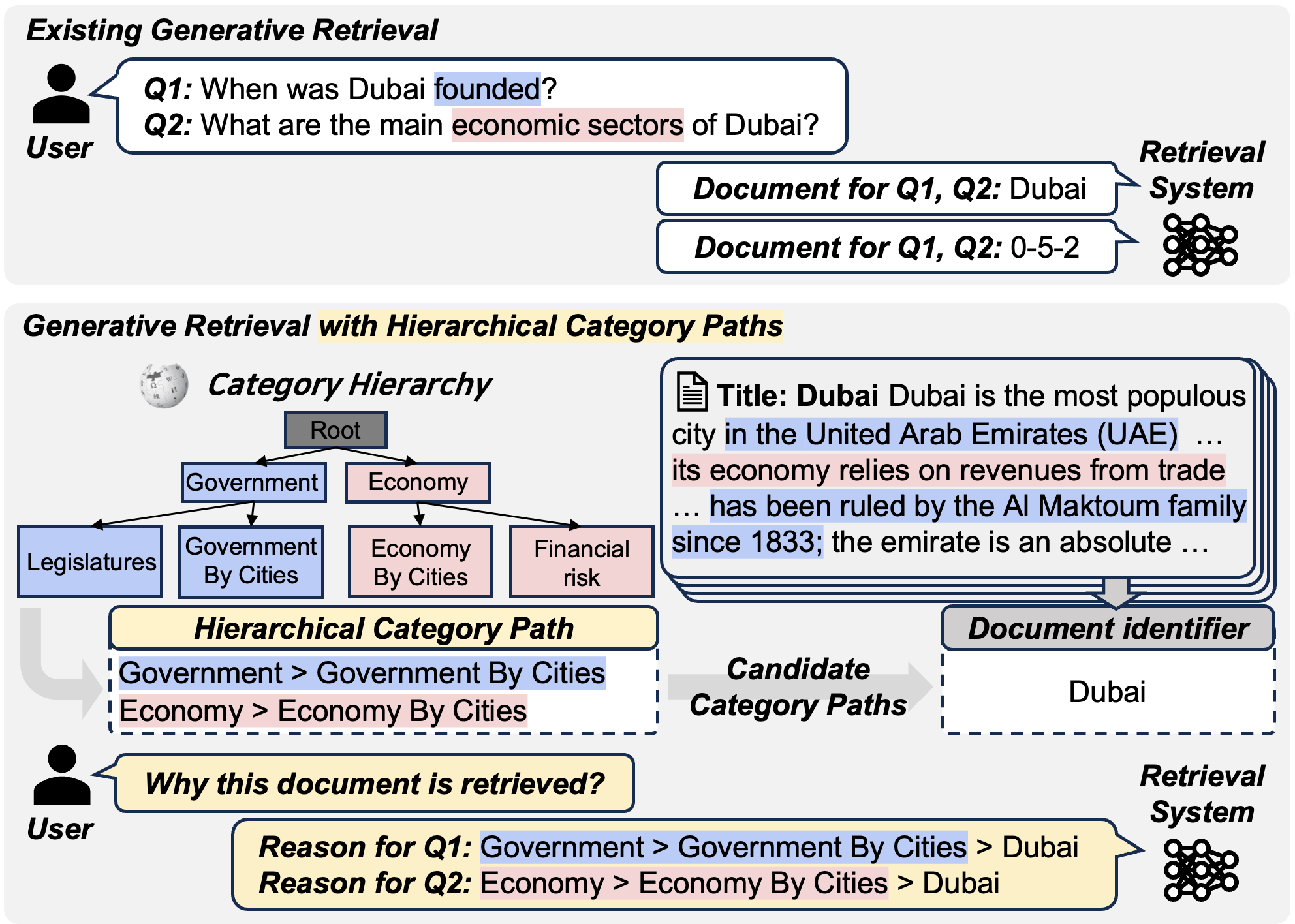}
    \caption{Existing \gr methods fail to explain why specific documents are retrieved, as they directly decode docid (Upper). In contrast, our \hype  provides clear explanations by generating query-related hierarchical category paths leading to the docid (Lower).}
    \label{fig:motivation}
\end{figure}
Recently, \textit{\gr} has emerged as a new paradigm in document retrieval.
It aims to directly decode document identifier (i.e., docid) for a given query by leveraging generative models such as BART~\cite{ACL/Liu2020/BART} and T5~\cite{JMLR/Raffel2020/T5}.
This paradigm enables end-to-end optimization of the retrieval process, allowing for fine-grained interaction between the input query and docid, and significantly reduces memory usage because it leverages only the parametric memory of a single generative model.
% This paradigm enables end-to-end optimization of retrieval with fine-grained query–docid interactions, while significantly reducing memory usage by relying solely on the parametric memory of a single generative model.

Even with these advantages, \gr continues to face the challenge of determining how to construct docids that effectively represent documents.
%As the docid serves as a representation of the entire document, defining one that accurately encapsulates the document's contents is both crucial and challenging. #==> rephrased to avoid an orphan sentence in this paragraph 
Defining a representative docid that accurately encapsulates the document's contents is crucial yet challenging.
Existing works on generative retrieval have categorized docid into two types: 
%\textbf{\sembasedocids} and \textbf{\textbasedocids}. 
\textit{semantic} docid and \textit{lexical} docid.
A semantic docid represents each document as a series of numbers (e.g., 0-5-2), where each number indicates a cluster index assigned over its dense representation.
This dense representation is encoded by a PLM-based encoder~\cite{NAACL/Devlin2019/BERT, JMLR/Raffel2020/T5} and clustered using methods such as hierarchical k-means~\cite{nips/Tay/DSI, nips/WangHWMWCXCZL0022/NCI} or product quantization~\cite{corr/abs-2208-09257/Ultron}.
% Semantic docid represents each document as a unique identifier, either a single number or a series of numbers (e.g., `0-5-2'), derived from the dense representations encoded by a encoder~\cite{NAACL/Devlin2019/BERT} and clustering techniques, such as hierarchical k-means clustering~\cite{nips/Tay/DSI, nips/WangHWMWCXCZL0022/NCI, corr/abs-2304-04171/GENRET} and product quantization~\cite{corr/abs-2208-09257/Ultron}. 
On the other hand, {\textbasedocid} represents each document as human-readable text, such as titles~\cite{iclr/CaoI0P21/GENRE}, keywords~\cite{SIGIR/Zhang2023/TSGen, CIKM/Wang2023/NOVO} and substring~\cite{nips/BevilacquaOLY0P22/SEAL}.

However, both existing approaches still lack explainability, which remains a significant limitation. 
For instance, in the upper part of Figure \ref{fig:motivation}, two types of queries related to the same document ``Dubai'', are presented. 
While the existing retrieval systems may return identifiers of relevant documents such as the lexical docid (i.e., Dubai) or semantic docid (i.e., 0-5-2), they fail to provide an explicit explanation that aligns with the different intention behind each query. 
\textbf{Specifically, they do not clarify the rationale behind retrieving a particular document for a specific query and fail to answer ``\textit{why is this document retrieved?}''.}
% Unlike \sembasedocid, \textbasedocid achieves interpretability by providing human-readable identifiers, but they still lack in \textit{explainability} regarding why these documents were retrieved. 
% In the upper part of Figure \ref{fig:motivation}, existing \gr methods {directly decode docid} when a query is given, making it impossible to provide users with explanations as an answer for \textit{``Why this document is retrieved?''}. 
% While users can recognize what a \textbasedocid represents, it is not always obvious why a certain docid is decoded for a specific query. 
The lack of explainability in retrieval is a critical issue, as it can undermine the reliability of retrieved documents and make it more difficult for users to explore additional information related to a specific query~\cite{anand2022explainable}.
To address this limitation, we aim to design a \gr framework that can provide retrieved document with clear and reasonable explanations for a  query. 

In this work, we propose \underline{\textbf{H}}ierarchical Categor\underline{\textbf{y}} \underline{\textbf{P}}ath-\underline{\textbf{E}}nhanced Generative Retrieval (\textbf{\hype}), which enhances explainability by generating hierarchical category paths step-by-step before decoding docids. 
Motivated by structured document categorization systems, such as Wikipedia category tree or Microsoft Academic taxonomy~\cite{shen-etal-2018-web}, \hype utilizes hierarchical category paths as explanations, progressing from broad to specific semantic categories. 
In the lower part of Figure \ref{fig:motivation}, when queries about document ``Dubai'' are given, \hype uses category paths like \textit{``Government $>$ Government by cities''} or \textit{``Economy $>$ Economy by cities''} to explain why document ``Dubai'' is retrieved for each query.
It 1) enables specific explanations for the document depending on the query by using hierarchical category paths that connect the query and the document, and 2) provides more reasonable and insightful explanation by reflecting the document's semantic structure through a coarse-to-fine manner. 
Additionally, \hype 3) provides users with clarity and guidance by providing explanation.

% Specifically, \hype consists of the following three steps: 1) constructing category paths based on an external high-quality semantic hierarchy and selecting appropriate candidate paths for each document using Large Language Models (\llm), 2) building a path-augmented dataset with candidate paths, and 3) optimizing a model with the path-augmented dataset. 
Specifically, \hype consists of the following three steps: 1) constructing category paths based on an external semantic hierarchy and selecting appropriate candidate paths for each document using \llm, 2) building a path-augmented dataset with candidate paths, and 3) optimizing a model with the path-augmented dataset. 
During inference phase, \hype conducts a pseudo-reasoning process\footnote{The rationale for this term is discussed in Appendix~\ref{appendix:pseudo_reasoning}.} by generating the hierarchical category path step-by-step to decode a docid, allowing it to serve as an explanation which enhances explainability.
Additionally, \hype employs \textit{\pars} strategy, which simultaneously considers multiple category paths for each query.
This strategy helps build a more robust retrieval system by capturing the semantic information of multiple category paths, thereby improving overall performance. 

Our experiments demonstrate that \hype not only improves the performance but also offers a high level of explainability, providing users with clarity and guidance.
Also, \hype can be applied orthogonally to various docid types, making it a versatile framework that can be seamlessly integrated into different generative retrieval systems.

We summarize our contributions as follows:
\begin{itemize}[leftmargin=*,topsep=2pt,itemsep=2pt,parsep=0pt]
    \item 
    We introduce \hype, an explainable generative retrieval framework that provides query-specific hierarchical category paths to explain document retrieval before decoding docids.
    
    \item 
    Our extensive experiments show that \hype consistently improves retrieval accuracy across diverse docid formats and retrieval corpus.

    \item 
    We show that the hierarchical category paths in \hype serve as effective retrieval explanations, helping users better understand and identify relevant documents in realistic search scenarios.
    
\end{itemize}

% appendix:pseudo_reasoning

\section{Related Work}
\label{sec:relwork}
% \paragraph{Document retrieval}
% Document retrieval is a representative task in information retrieval (IR) that involves searching for documents relevant to a given query within a document corpus. Traditionally, it has been accomplished through \textit{sparse retrieval} and \textit{dense retrieval}.  
% Sparse retrieval~\cite{robertson09/bm25, sigir/FormalPC21/SPLADE} employs term-matching scoring methods, such as BM25~\cite{robertson09/bm25}, to calculate relevance scores for retrieving documents.
% However, they struggle with the issue of lexical mismatch~\cite{Lin2020PretrainedTF} due to terms such as homonyms and synonyms. 
% % To address these limitations, 
% Dense retrieval ~\cite{EMNLP/KarpukhinOMLWEC20/DPR, sigir/KhattabZ20/ColBERT, iclr/XiongXLTLBAO21/ANCE} has emerged with the recent development of pre-trained language models (PLMs).
% This approach utilizes encoder-based \plm, such as BERT~\cite{NAACL/Devlin2019/BERT}, to obtain dense representations for both queries and documents, and retrieve relevant documents through an Approximate Nearest Neighbor (ANN) index~\cite{Malkov2016/EfficientANN}.
% Although dense retrieval methods have demonstrated impressive performance by capturing deep semantic information that sparse retrieval methods cannot, it still has the limitation of not being able to optimize the model in an end-to-end manner.
% To address this limitations, \gr has emerged as an alternative to traditional document retrieval methods.

\paragraph{Explainable retrieval system.}
Explainable retrieval has relied on feature attribution methods~\cite{robertson09/bm25, Verma2019LIRME}, which aim to explain relevance by quantifying the importance of individual input terms to the retrieval decisions.
However, they are inherently limited to superficial keyword lists, often failing to provide sufficient semantic context or capture the underlying reasoning behind retrieval decisions~\cite{anand2022explainable}.
To overcome this limitation, free text explanation methods~\cite{Liu2024RaCTRC, Pulakurthi2025XCoTET} have emerged, leveraging \llm to provide detailed natural language explanation. 
However, the substantial cost of generating natural language explanations results in high latency~\cite{jia2025improving}, rendering them impractical for real-time retrieval.
In this work, we utilize hierarchical category paths to address these challenges, sufficiently capturing the retrieval reasoning process while mitigating the inference latency associated with free-text generation.

% One line of work in explainable retrieval focuses post-hoc explanation methods~\cite{Chowdhury2022RankLIMELM, Verma2019LIRME}.
% Specifically, they employ external models to explain the behavior of the retrieval system.
% However, since these methods rely on an external model rather than the retrieval model itself, they often fail to fully capture the retriever's decision mechanism~\cite{Rudin2018StopEB}.
% To address this limitation, explainable-by-architecture models have been proposed, aiming to construct models that are inherently explainable~\cite{sigir/KhattabZ20/ColBERT, Lucchese2022ILMARTIR}. 
% These methods incorporate explainability directly into the model architecture, ensuring that the retrieval process itself serves as the explanation~\cite{anand2022explainable}.

% \subsection{Generative Retrieval}
% \label{generative_retrieval}
\paragraph{Generative retrieval.}
% Unlike traditional retrieval methods, \gr leverages a single pre-trained generative model to directly decode docids relevant to a query, enabling end-to-end optimization of the retrieval process~\cite{nips/Tay/DSI}.
% Additionally, this approach reduces reliance on external indexing, lowering the system’s demand for storage resources~\cite{nips/Tay/DSI, SIGKDD/Tang2023/SE-DSI}. 
Generative retrieval uses a single generative model to directly decode docids for a query, enabling end-to-end optimization while reducing reliance on external indexing and storage resources~\cite{nips/Tay/DSI}.
% Despite these advantages, explainability at the level of the retrieval process remains limited
Despite these advantages, the explainability of generative retrieval remains underexplored.
Many existing methods represent docids as series of numbers~\cite{nips/WangHWMWCXCZL0022/NCI, corr/abs-2304-04171/GENRET, corr/abs-2208-09257/Ultron}, which makes it difficult for users to understand the retrieval system’s decisions.
Although recent methods decode human-readable docids to improve interpretability~\cite{Lee2023GLENGR, CIKM/Wang2023/NOVO, SIGIR/Zhang2023/TSGen}, they do not provide explanations of why the document is retrieved in response to a query or how the retrieval decision is made.
To address this gap, we propose a \gr framework that decode docids while also providing explanations of the retrieval process.

\section{Preliminaries}
\label{sec:preliminary}
In this section, we explain the main components and the overall process of generative retrieval.

\subsection{Task Formulation}
\label{subsec:genretrievaltask}
Given a corpus $\mathcal{C} = \{D_{1}, D_{2}, \dots , D_{n}\}$ where $D$ represents a document, generative retrieval aims to autoregressively generate the document identifier (i.e. docid) of the relevant document for a given query. 
To this end, the model is optimized for \textbf{indexing task} and \textbf{retrieval task}. 
The indexing task involves taking a document as the input and decoding the corresponding docid, described by:
\begin{equation}
    \label{eq:indexinglEq}
    \mathcal{M}^{\theta}(d \mid D) = \prod_{t=1}^{n} \mathcal{M}^{\theta}(d_t \mid D, d_{<t}),
\end{equation}
where $\mathcal{M}^{\theta}$ is a generative model, $D$ is a document, $d$ is the target docid, and $n$ is the token length of the target docid. 
The retrieval task processes a query as the input and decoding the relevant docid:
\begin{equation}
    \label{eq:retrievalEq}
    \mathcal{M}^{\theta}(d \mid q) = \prod_{t=1}^{n} \mathcal{M}^{\theta}(d_t \mid q, d_{<t}),
\end{equation}
where $q$ is a query.
In performing the aforementioned two tasks, it is crucial to address two key aspects:  1) effectively represent the long document $D$ and 2) construct the docid $d$ that captures the overall semantic information of the $D$.

% While recent works incorporate additional optimizations, such as ~\cite{corr/abs-2304-04171/GENRET} and ~\cite{Lee2023GLENGR}, specifically designed for each method, this work does not address those approaches.
% At inference phase, the model generates a ranked list of docid $d$ for relevant documents as the retrieval result for the given query.
During inference, given an input query $q$, the model produces a top-$K$ ranked list of docids that have the largest likelihoods $\mathcal{M}^{\theta}(d \mid q)$. 
To ensure the generation of valid docids, the model employs constrained decoding, which mostly uses constrained beam search with prefix trie~\cite{iclr/CaoI0P21/GENRE} or FM-index~\cite{nips/BevilacquaOLY0P22/SEAL}.

\begin{figure*}[t]
    \centering
    \includegraphics[width=\textwidth]{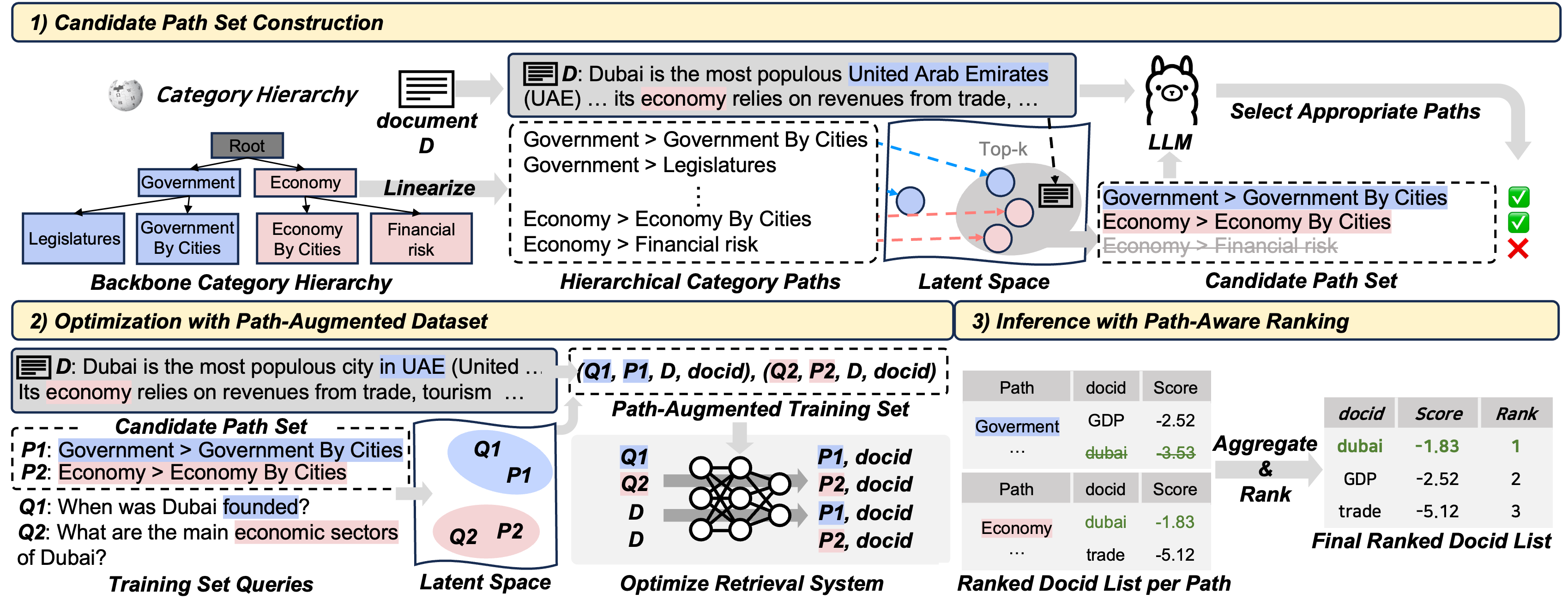}
    \caption{Overview of \hype framework. (1) \hype constructs category paths using an external semantic hierarchy and employs LLM to select appropriate candidate paths for each document. (2) It then links queries to semantically relevant paths to form a path-augmented training set for retrieval optimization. (3) During inference, \hype employs path-aware ranking strategy to determine the final docid ranking by considering multiple paths.}
    \label{fig:method}
\end{figure*}

\subsection{Document Representation and Identifier}
\label{subsec:docrepdocid}
% \subsubsection{Document representation}
% \label{subsubsec:docrep}
% As previously mentioned, in 
% \textbf{Document representation.}
\paragraph{Document representation.}
% For the indexing task, each document is used as the input, making it crucial to define effective input representations of the long document content while preserving as much of its information as possible within the context length of the language model. 
For the indexing task, each document is used as the input.
This makes it crucial to define effective input representations of the long document while preserving as much of its information as possible within the context length of the language model.
The primary approaches to effectively representing documents are FirstP~\cite{nips/Tay/DSI} and DaQ~\cite{nips/WangHWMWCXCZL0022/NCI}. 
FirstP uses only the first $k$ tokens from the document as input representation, while DaQ randomly extracts chunks from the document. 
% \subsubsection{Document identifier}
% \label{subsubsec:docid}

\paragraph{Document identifier.}
% \noindent\textbf{Document identifier.}
To ensure that docid effectively encodes information of document, a variety of methods have been proposed. 
% It can be broadly categorized into \textit{\sembasedocid} and \textit{\textbasedocid}.
% Semantic docid represents each document as a series of numbers, where each number corresponds to a cluster index derived from the document’s dense representation.
% It is composed of dense representations encoded by a PLM-based encoder~\cite{NAACL/Devlin2019/BERT}, which are mapped to discrete cluster indices.
Semantic docid represents each document as a series of numbers, where each number corresponds to a cluster index derived from the document’s dense representation. 
This representation is encoded by a PLM~\cite{NAACL/Devlin2019/BERT} and mapped to discrete cluster indices using methods such as hierarchical k-means~\cite{nips/Tay/DSI, nips/WangHWMWCXCZL0022/NCI} or product quantization~\cite{corr/abs-2208-09257/Ultron}.
Lexical docid is a textual format designed to effectively convey the semantic content of a document. 
It can be constructed using various forms, such as title~\cite{iclr/CaoI0P21/GENRE}, substrings~\cite{nips/BevilacquaOLY0P22/SEAL}, keywords~\cite{SIGIR/Zhang2023/TSGen, Lee2023GLENGR}, URL~\cite{corr/abs-2208-09257/Ultron}, and pseudo query~\cite{SIGKDD/Tang2023/SE-DSI}. 
% Title and URL are used as docid directly from the dataset. 
% Substrings are specific n-grams within the document. 
% Keywords are extracted using methods such as BM25~\cite{robertson09/bm25}, or pre-trained language models (PLMs).
% Pseudo query is generated using query generation models, such as docT5query~\cite{NogueiraL19/docT5query}.

\subsection{Optimization}
\label{subsec:traingr}
% \subsubsection{Optimization via multi-task learning}
% \label{subsec:optviamtl}
% \textbf{Optimization via multi-task learning.}
\paragraph{Optimization via multi-task learning.}
Given a training dataset that consists of (query, document, docid), denoted by $\mathcal{X} = \{(q, D, d)\}$, the model is trained for both the indexing and retrieval tasks, maximizing the likelihoods in \eqref{eq:indexinglEq} and~\eqref{eq:retrievalEq}:
\begin{equation}
\label{eq:optimization}
\max_\theta \sum_{(q, D, d) \in \mathcal{X}}  \mathcal{M}^\theta(d \mid D) + \mathcal{M}^\theta(d \mid q)
\end{equation}
% \subsubsection{Indexing with synthetic query}
% \label{subsec:indexwithsynthetic}
\paragraph{Indexing with synthetic query.}
In indexing task, documents are long and contain extensive information;
however, in retrieval task, queries are relatively short and request specific information.
To bridge this discrepancy, recent studies~\cite{corr/abs-2206-10128/DSI-QG, nips/WangHWMWCXCZL0022/NCI, corr/abs-2304-04171/GENRET} have tried to integrate synthetic queries, generated by query generation models~\cite{NogueiraL19/docT5query}, into the training phase. 
The synthetic queries improve the retrieval performance of generative retrieval models by effectively reducing the gap between queries and documents. 
Note that these queries are treated as alternative document representation, and are used as input for the indexing task~\cite{corr/abs-2206-10128/DSI-QG, corr/abs-2304-04171/GENRET}.

\section{Proposed Method}
\label{sec:method}
In this section, we present \underline{\textbf{H}}ierarchical categor\underline{\textbf{y}} \underline{\textbf{P}}ath-\underline{\textbf{E}}nhanced generative retrieval (\textbf{\hype}).
% , which improves explainability by generating hierarchical category paths step-by-step before decoding docid. 
% Our hierarchical category paths serve as pseudo-reasoning process, providing a reasonable and reliable explanation for why this document is retrieved. 
% Furthermore, \hype employs \textit{\pars} strategy, which simultaneously considers multiple pseudo-reasoning paths to determine the documents to retrieve. 
The overall framework is shown in Figure~\ref{fig:method}.

% 3.1
\subsection{Candidate Path Set Construction}
\label{subsec:stepone}
% \subsubsection{Hierarchical category path} 
% The first step is to assign each document into multiple hierarchical category paths, by matching their semantic relevance. 
The first step of our \hype framework is to construct a set of candidate hierarchical category paths for each document. 
To ensure explainability, these paths should satisfy the following criteria: \textit{Semantic Hierarchy}, \textit{Generalizability}, and \textit{Specificity} (see Appendix~\ref{appendix:backbone_category hierarchy} for details).
To achieve this, we first construct the high-quality backbone hierarchy for category paths. 
Then, for each document, we (1) filter out category paths based on semantic similarity calculated by a pre-trained text encoder, and (2) select several category paths that comprehensively represent the content of the document while specifically addressing certain topics within the document by the help of reasoning capabilities of \llm.

\paragraph{Hierarchical category path collection.} 
\label{subsubsec:pathcollection}
In the open-domain retrieval task, the category (or topic) hierarchy must encompass both a broad range of domain categories (i.e. width of tree) and sufficient semantic granularity (i.e. depth of tree) to ensure comprehensive and accurate retrieval system.
% To this end, we leverage Wikipedia's category tree\footnote{\url{https://en.wikipedia.org/wiki/Special:CategoryTree}} as our backbone hierarchy of categories, setting the \textit{Main Topic classification} category\footnote{\url{https://en.wikipedia.org/wiki/Category:Main_topic_classifications}} as the root node of the hierarchy.
To this end, we leverage Wikipedia's category tree as our backbone hierarchy of categories, setting the \textit{Main Topic classification} category as the root node of the hierarchy.
This hierarchy is specifically designed to systematically categorize ``real-world wikipedia documents'', which cover a wide range of domains and provide specific and detailed semantic information.
% Its broad and deep structure ensures that the hierarchy can effectively meet the demands of open-domain tasks by capturing diverse domains and providing sufficient granularity. 
Considering the vast and complex nature of Wikipedia's category tree, we limit the scraping process to a depth of four to construct our backbone hierarchy. 
Then, we linearize all the paths within the hierarchy and convert them into a sequence of strings, thereby enabling more efficient processing and manipulation. 
The entire set of linearized category paths is denoted by $\mathcal{P}$.
While we adopt Wikipedia's category tree as our backbone hierarchy for its broad applicability to open-domain document retrieval, \hype is not restricted to it.
For specialized domains, domain-specific taxonomies can be readily integrated in a plug-and-play fashion~\cite{Kang2024TaxonomyguidedSI}.
We further discuss this point in Appendix~\ref{appendix:backbone_category hierarchy}.
% Discussion of the backbone hierarchy and statistics of category paths are presented in Appendix~\ref{appendix:backbone_category hierarchy}.

\paragraph{Candidate path set construction.} 
\label{subsubsec:path_assignment}
% Subsequently, we leverage the knowledge of an LLM to assign appropriate category paths to each document in the corpus.
% However, due to the limited context length of LLMs, it is infeasible to provide all possible paths in the category hierarchy (collected in Section~\ref{subsubsec:pathcollection}) as input.
% To address this issue, we first filter the candidate path set for each document $D$ using a bi-encoder.
% Specifically, we construct a pre-candidate path set $\hat{\mathcal{P}}_{D}$ by selecting the top-$k$ most similar paths from the full path set $\mathcal{P}$, defined as
% $\hat{\mathcal{P}}_{D} = \operatorname*{argTop-}k_{p \in \mathcal{P}} \, \text{sim}(E(D), E(p))$,
% where $E(\cdot)$ denotes the encoder, $\text{sim}(\cdot)$ is the cosine similarity, and $k$ is the number of pre-candidate paths per document.
Subsequently, we utilize the LLM to assign appropriate category paths to each document within the corpus.
However, due to the context length of LLM, it is impossible to input all possible paths within the category hierarchy.
% (collected in Section~\ref{subsubsec:pathcollection}).
Thus, we first filter out path set for each document $D$ by leveraging a bi-encoder.
The pre-candidate path set $\hat{\mathcal{P}}_{D}$ is obtained as follows:
\begin{equation}
    \label{eq:selectcandpathset}
    \hat{\mathcal{P}}_{D} = \operatorname*{argTop-}_{p \in \mathcal{P}}\hspace{-1pt}k\text{ sim}(E(D), E(p)),
\end{equation}
where $E(\cdot)$ is the encoder, $\text{sim}(\cdot)$ is a cosine similarity, and $k$ is the number of pre-candidate paths for each document. 
Then, given the document $D$ and its pre-candidate path set $\hat{\mathcal{P}}_{D}$, we leverage LLM to generate the final path set $\mathcal{P}_{D}$, selecting up to three paths that best represent the document.
For more details, please refer to Appendix~\ref{appendix:detailtrainingset}
% \footnote{We use Llama-3-8B-Instruct as \llm.}
% \selectFinalPathSetFull
% \begin{equation}
% \label{eq:selectfinalpathset}
%     \mathcal{P}_{D} = \text{\llm}({D}, \hat{\mathcal{P}}_{D}).
% \end{equation}
% $\mathcal{P}_{D}$ refers to the final candidate path set of the document $D$. 
% \subsubsection{Assign paths to document.} 
% \label{subsubsec:assignpathstodocument}

% \subsection{Build Path Augmented training set}

\subsection{Optimization with Category Path}
\label{subsec:steptwo}
The second step is to augment the training set $\mathcal{X}$ with path, building a path-augmented training set $\mathcal{X}^{+} = \{ (q, p^q, D, d) \}$. 
To achieve this, we first (1) link each query to one of the document's candidate paths based on semantic similarity computed by pre-trained encoder, and then (2) utilize the resulting query-path pairs together with the document-path pairs to optimize the
retrieval model.

\paragraph{Linking path with query.} 
\label{subsubsec:assigndocumentspathtoquery}
Using the candidate path set for each document, we build a \textit{path augmented training set} $\mathcal{X}^{+}$. 
For each query-document pair in the training set $(q,D,d)\in\mathcal{X}$, we link the query $q$ to its most relevant path among the paths in the document's candidate path set ${P}_{D}$. 
This linking can be described as follows:
% \assignQueryPath
\begin{equation}
\label{eq:assignquerypath}
p^{q} = \operatorname*{argmax}_{p\in \mathcal{P}_{D}} \text{ sim}(E(q), E(p)),    
\end{equation}
where $p^{q}$ is the path linked to the query $q$. 
This process is then applied to all queries in the training set. 
In the end, we construct the path-augmented training set, denoted by $\mathcal{X}^{+} = \{ (q, p^q, D, d) \}$.

\paragraph{Optimization.} 
\label{subsubsec:modeloptimization}
By leveraging the path-augmented training set $\mathcal{X}^{+}$, we train our model $\mathcal{M}^{\theta}$ on both indexing and retrieval tasks, as described in \ref{subsec:genretrievaltask}. 
Our optimization follows the same strategy as standard generative retrieval in \ref{subsec:genretrievaltask}, with the only difference being the addition of path information as follows:
\begin{equation}
\label{eq:optimization}
% \max_\theta \sum_{(q, p^q, D, d) \in \mathcal{X}^{+}}  \mathcal{M}^\theta(p^{q},d \mid D) + \mathcal{M}^\theta(p^{q}, d \mid q)
\max_\theta \sum \mathcal{M}^\theta(p^{q},d \mid D) + \mathcal{M}^\theta(p^{q}, d \mid q)
\end{equation}
% That is, the model is optimized by indexing loss and retrieval loss. 
% First, the indexing loss is defined by
% \begin{equation}
% \label{eq:indexinglossterm}
%     \mathcal{L}_{\text{indexing}} = - \sum_{p\in\mathcal{P}_{D}}\log \mathcal{M}^{\theta}(p, d \mid D)
% \end{equation}
% where $p$ is a path from $D$'s candidate path set. 
% Note that there are multiple paths assigned to the document, so we optimize our model by using all candidate paths in $\mathcal{P}_D$. 
% % Through this process, the model can effectively memorize the content of the documents while mapping them to their relevant paths. 
% Next, the retrieval loss is denoted as follows:
% \begin{equation}
% \label{eq:retrievallossterm}
%     \mathcal{L}_{\text{retrieval}} = - \log \mathcal{M}^{\theta}(p, d \mid q)
% \end{equation}
% Finally, we define our total loss as the sum of the indexing loss and the retrieval loss, regarding the training process as a multi-task learning.

\subsection{Inference with Path-Aware Ranking} 
\label{subsec:stepthree}
During inference, \hype generates the final ranked list of docids through two stages: 1) \textit{path generation stage} and 2) \textit{docid decoding stage}. 
First, in the path generation stage, our model $\mathcal{M}^{\theta}$ generates up to $K_p$ hierarchical category paths, each of which is denoted by $p_j$ for $j=1,\dots,K_p$, by using beam search; 
these are query-specific hierarchical category paths that encapsulate various topics related to the given query.
Next, in the docid decoding stage, the model uses each generated hierarchical category path as the decoder's input context and then applies constrained beam search to decode $m$ docids.
For each path $p_j$, the model outputs $m$ number of docid-score pairs as follows:
\begin{equation}
\label{eq:construct_pair_aggregating}
Y_j = \{(d_i, s_i)\sim\ \mathcal{M}^\theta(\cdot\mid q,p_j)\}_{i=1}^{m},
\end{equation}
where $s_i$ represents the score for the docid $d_i$ conditioned on the category path $p_j$.
The remaining process is to aggregate $K_p$ number of docid-score pair sets for making the final ranked list of docids.
% \begin{equation}
% \label{eq:construct_pair_aggregating_2}
% Y' = \bigcup_{j=1}^{K_p} Y_j
% \end{equation}
% where $Y'$ is the union of all the docid-score pairs generated from the $K_p$ paths.
At this point, we remain only unique docid with the highest score, resulting in $\tilde{Y}$. 
\begin{equation}
\label{eq:reranking}
\tilde{Y} = \left\{(d, s) \mid s = \max\{ s' | (d, s')\in Y_j \}, \forall (d, s) \in \cup_{j=1}^{K_p} Y_j \right\}
\end{equation}
From the set of unique docid-score pairs, we obtain the final ranked list by sorting their scores in descending order, $Y_{\text{final}} = \text{sort}(\tilde{Y})$.
By utilizing \textit{\pars} strategy, \hype can effectively capture the semantic information of an input query from multiple hierarchical category paths.

\begin{table*}[t]
\centering
\small
\resizebox{.99\textwidth}{!}{
\label{tab:nq320k}
\begin{tabular}{ccccccccccccccccc}
\toprule
\multirow{2.5}{*}{\textbf{Method}} 
& \multicolumn{4}{c}{\textbf{Full test}} 
& \multicolumn{4}{c}{\textbf{Seen test}}
& \multicolumn{4}{c}{\textbf{Unseen test}}
\\
\cmidrule(lr){2-5} \cmidrule(lr){6-9} \cmidrule(lr){10-13}
& R@1 & R@10 & R@100 & M@100 & R@1 & R@10 & R@100 & M@100 & R@1 & R@10 & R@100 & M@100 \\

\midrule

% GENRE~\cite{iclr/CaoI0P21/GENRE}
% & 55.2 & 67.3 & 75.4 & 59.9
% & 69.5 & 83.7 & 90.4 & 75.0
% & 6.0 & 10.4 & 23.4 & 7.8
% \\
% \cdashline{1-13}

Title docid
& 62.2 & 78.7 & 89.3 & 68.6
& 64.8 & 81.5 & 90.1 & 71.2
& 53.1 & 68.9 & 80.4 & 59.3
\\

+ \hype 
& 63.6$^*$ & 83.5$^*$ & 90.1$^*$ & 71.0$^*$
& 66.4$^*$ & 86.3$^*$ & 92.6$^*$ & 73.9$^*$
& 53.7$^*$ & 73.6$^*$ & 81.7$^*$ & 61.0$^*$
\\

\rowcolor{gray!20}
\textbf{Improvement}
& \textbf{+2.3\%} & \textbf{+6.1\%} & \textbf{+2.5\%} & \textbf{+3.5\%} 
& \textbf{+2.5\%} & \textbf{+5.9\%} & \textbf{+2.8\%} & \textbf{+3.8\%}
& \textbf{+1.1\%} & \textbf{+6.8\%} & \textbf{+1.6\%} & \textbf{+2.9\%} 
\\
\midrule

% NOVO~\cite{CIKM/Wang2023/NOVO}
% & 69.3 & 89.7 & 95.9 & 76.7
% & - & - & - & -
% & - & - & - & -
% \\
% \cdashline{1-13}

Keyword docid
& 61.8 & 77.1 & 85.5 & 67.6
& 67.3 & 82.3 & 89.9 & 73.0
& 43.0 & 59.0 & 70.4 & 48.8
\\

+ \hype
& 60.7 & 79.1$^*$ & 86.2$^*$ & 67.6
& 66.6 & 84.6$^*$ & 90.7$^*$ & 73.4$^*$
& 40.1 & 60.2$^*$ & 70.6$^*$ & 47.5
\\

% \cdashline{2-13}
\rowcolor{gray!20}
\textbf{Improvement}
& -1.8\% & \textbf{+2.6\%} & \textbf{+0.8\%} & +0.0\%
& -1.0\% & \textbf{+2.8\%} & \textbf{+0.9\%} & \textbf{+0.5\%}
& -6.7\% & \textbf{+2.0\%} & \textbf{+0.3\%} & -2.7\%
\\
\midrule

% SEAL~\cite{nips/BevilacquaOLY0P22/SEAL}
% & 59.9 & 81.2 & 90.9 & 67.7
% & - & - & - & -
% & - & - & - & -
% \\
% \cdashline{1-13}

Summary docid
& 60.9 & 78.8 & 84.1 & 67.6
& 65.7 & 84.1 & 88.6 & 72.6
& 44.0 & 60.5 & 68.5 & 50.1
\\

+ \hype
& 61.5$^*$ & 79.6$^*$ & 85.2$^*$ & 68.3$^*$
& 66.3$^*$ & 84.6$^*$ & 89.8$^*$ & 73.2$^*$
& 44.8$^*$ & 62.2$^*$ & 69.4$^*$ & 51.3$^*$
\\
% \cdashline{2-13}
\rowcolor{gray!20}
\textbf{Improvement}
& \textbf{+1.0\%} & \textbf{+1.0\%} & \textbf{+1.3\%} & \textbf{+1.0\%} 
& \textbf{+0.9\%} & \textbf{+0.6\%} & \textbf{+1.4\%} & \textbf{+0.8\%}
& \textbf{+1.8\%} & \textbf{+2.8\%} & \textbf{+1.3\%} & \textbf{+2.4\%}
\\
\midrule

% GENRET~\cite{corr/abs-2304-04171/GENRET}
% & 68.1 & 88.8 & 95.2 & 75.9
% & 70.2 & 90.3 & 96.0 & 77.7
% & 62.5 & 83.6 & 92.5 & 70.4
% \\
% \cdashline{1-13}

Atomic docid
& 65.3 & 83.5 & 89.3 & 72.2
& 70.2 & 88.3 & 93.5 & 77.2
& 48.6 & 66.8 & 74.9 & 55.0
\\

+ \hype
& 64.5 & 84.2$^*$ & 90.2$^*$ & 71.9
& 69.5 & 88.6$^*$ & 93.8$^*$ & 76.8
& 47.2 & 68.7$^*$ & 77.6$^*$ & 55.0
\\
% \cdashline{2-13}
\rowcolor{gray!20}
\textbf{Improvement}
& -1.2\% & \textbf{+0.8\%} & \textbf{+1.0\%} & -0.4\% 
& -1.0\% & \textbf{+0.3\%} & \textbf{+0.3\%} & -0.5\%
& -2.9\% & \textbf{+2.8\%} & \textbf{+3.6\%} & +0.0\%
\\
\bottomrule
\end{tabular}
}
\caption{Retrieval accuracy of baselines and our \hype framework on the \nq. $*$ denotes the statistical significance on paired t-test $p<0.05$.}
\label{tab:nq320k}
\end{table*}
\begin{table}[t]
\centering
\small
\resizebox{.99\columnwidth}{!}{
\begin{tabular}{ccccc}
\toprule
\multirow{1}{*}{\textbf{Method}}  
& R@1 & R@10 & R@100 & M@10 \\

\midrule

% NOVO~\cite{CIKM/Wang2023/NOVO}
% & 49.1 & 80.8 & 92.5 & 59.2

% \\

% \cdashline{1-5}

Keyword docid
& 31.7 & 61.2 & 77.2 & 41.0 \\

+ \hype
& 32.2$^*$ & 62.7$^*$ & 78.5$^*$ & 41.9$^*$ \\

\rowcolor{gray!20}
\textbf{Improvement}
& \textbf{+1.6\%} & \textbf{+2.5\%} & \textbf{+1.7\%} & \textbf{+2.2\%} \\

\midrule

% SEAL~\cite{nips/BevilacquaOLY0P22/SEAL}
% & 25.9 & 68.6 & 87.9 & 40.2

% \\

% \cdashline{1-5}

Summary docid
& 28.1 & 55.5 & 71.5 & 36.8 \\

+ \hype
& 28.4$^*$ & 57.5$^*$ & 73.1$^*$ & 37.8$^*$ \\

\rowcolor{gray!20}
\textbf{Improvement}
& \textbf{+1.1\%} & \textbf{+3.6\%} & \textbf{+2.2\%} & \textbf{+2.7\%} \\

\midrule

% GENRET~\cite{corr/abs-2304-04171/GENRET}
% & 47.9 & 80.8 & 91.6 & 59.2

% \\
% \cdashline{1-5}

Atomic docid
& 43.9 & 73.6 & 85.6 & 53.8 \\

+ \hype
& 44.9$^*$ & 74.6$^*$ & 87.1$^*$ & 54.7$^*$ \\

\rowcolor{gray!20}
\textbf{Improvement}
& \textbf{+2.3\%} & \textbf{+1.4\%} & \textbf{+1.8\%} & \textbf{+1.7\%} \\

\bottomrule
\end{tabular}
}
\caption{Retrieval accuracy on the \msmarco. $*$ denotes statistical significance on paired t-test $p<0.05$.}
% \caption{Retrieval accuracy  of baselines and our \hype on the \msmarco. $*$ denotes the statistical significance on paired t-test $p<0.05$.}
\label{tab:msmarco}
\end{table}

\section{Experiments}
\label{sec:exp}

In this section, we design and conduct our experiments to answer the following research questions:
\begin{itemize}[leftmargin=*,topsep=2pt,itemsep=2pt,parsep=0pt]
    % \item \textbf{RQ1:} Can hierarchical category path serve as effective explanation for document retrieval tasks? 
    % \item \textbf{RQ2:} Can HyPE provide more effective and reasonable explanations compared to other methods for the user’s query?
    \item \textbf{RQ1:} Can \hype improve retrieval accuracy?
    \item \textbf{RQ2:} Can hierarchical category paths in \hype serve as effective explanations for retrieval?
    \item \textbf{RQ3:} Can explanations of \hype help real-world users in search systems?
    % \item \textbf{RQ4:} Can \textit{\pars} strategy of \hype improve retrieval accuracy?
\end{itemize}

\subsection{Experimental Settings}
\label{subsec:experimentsetup}
\paragraph{Dataset.}
\label{subsec:dataset}
We conduct our experiments on two datasets, \textbf{\nq}~\cite{TACL/KwiatkowskiPRCP19/NQ} and \textbf{\msmarco}~\cite{NeurIPS/Nguyen2016/msmarco}, which have been widely utilized in previous works~\cite{nips/Tay/DSI, nips/WangHWMWCXCZL0022/NCI}. 
% \nq is a dataset derived from Google’s Natural Questions (NQ) and comprises 307,373 query-document pairs along with 7,830 test queries, all linked to 109,739 documents sourced from Wikipedia.
For \nq, we divide the test set into two subsets, \textit{seen} and \textit{unseen}, following the setup in ~\cite{nips/WangHWMWCXCZL0022/NCI, corr/abs-2304-04171/GENRET}, where the \textit{seen} test includes queries whose annotated target documents are present in the training set, and the \textit{unseen} test consists of queries with no labeled documents in the training set.
% On the other hand, \msmarco comprises queries and web pages sourced from Bing search. 
% We use the \msmarco document ranking dataset, which contains 366,235 query-document relevant pairs and 5,187 test queries with 323,569 documents.
More details are provided in Appendix~\ref{appendix:dataset_overview}.

\paragraph{Evaluation metrics.}
\label{subsec:metrics}
We report Recall and Mean Reciprocal Rank (MRR) for \nq and \msmarco. 
For \nq, we use Recall@\{1, 10, 100\} and MRR@100. 
For \msmarco, we use Recall@\{1, 10, 100\} and MRR@10 as done in previous works~\cite{corr/abs-2304-04171/GENRET, CIKM/Wang2023/NOVO}.

\paragraph{Baselines.}
To validate the effectiveness of \hype across diverse \gr settings, we conduct experiments on four representative docid types, introduced in Section \ref{subsec:docrepdocid}, as our baseline. 

% In this paper, the docid types we used are as follows:
\begin{itemize}[leftmargin=*,topsep=2pt,itemsep=2pt,parsep=0pt]
    \item 
    \textbf{Title docid} uses a document's title as docid. 
    For documents without a title, we use the first 16 tokens of the document as a title, following the approach used in \cite{corr/abs-2304-04171/GENRET}. 
    
    \item 
    \textbf{Keyword docid} uses a sequence of keywords as docid that effectively represent the document.
    For \nq, we use 3 keywords, while for \msmarco, we extract 5 keywords.
    
    \item 
    \textbf{Summary docid} uses the document summary as docid.
    Although it has not been attempted before, a similar structure using substrings is employed in \cite{nips/BevilacquaOLY0P22/SEAL}.
    
    \item 
    \textbf{Atomic docid} uses a unique arbitrary integer as docid. 
    We assign each document an integer and generates a corresponding new token for it.
\end{itemize}

We intentionally do not consider semantic docids (+\hype) in our experiments. This is because semantic docids are constructed based on techniques such as hierarchical clustering, and thus inherently embed a semantic structure. Given that these structures are already formed in a coarse-to-fine manner, prepending hierarchical category paths to them can contradict the coarse-to-fine principle.

Furthermore, existing generative methods employ various  architectures and optimization techniques, which may introduce additional factors affecting performance. 
\textbf{To specifically assess the impact of \hype, we adopt the basic form of generative retrieval described in Section \ref{sec:preliminary} as our baseline.}
% To isolate and specifically assess the impact of \hype, we use the basic form of generative retrieval introduced in Section \ref{sec:preliminary} as our main baseline. 
This approach ensures a direct comparison between plain docids and those enhanced with \hype, isolating the effects of \hype itself from other architectural or optimization differences. 
For more details, please refer to the Appendix~\ref{appendix:Implementationdetail}.
% In addition, we also report the performance of existing methods as a reference for existing generative methods. 
% % Furthermore, we additionally report the performance of existing methods to provide a reference for the performance of existing generative methods. 
% Specifically, we set a \textit{reference baseline} for each docid type by selecting the method that achieves the highest performance on the same dataset as ours.
% The detailed criteria used for selecting reference baselines are provided in \ref{sec:dataset_overview}. 

\paragraph{Implementation details.}
% We use T5-base~\cite{JMLR/Raffel2020/T5} as our backbone model. 
% For the input of the indexing task, we utilize the FirstP approach as our document representations and five synthetic queries. (Section~\ref{subsec:docrepdocid}).
% During the inference of \hype, we generate three category paths (i.e., $K_p=3$), and for the docid decoding stage, we use constrained beam search with a beam size of 100 (i.e., $m=100$). 
We use T5-base~\cite{JMLR/Raffel2020/T5} as our backbone model. 
For the input of the indexing task, we utilize the FirstP and use additional five synthetic queries.
During the inference of \hype, we generate three category paths (i.e., $K_p=3$), and use constrained beam search with a beam size of 100 (i.e., $m=100$). 
% More details about this part are provided in Appendix~\ref{appendix:Implementationdetail}.

% \input{tables/STS_result}
% \begin{figure}[t]
%     \centering
%     \includegraphics[width=0.99\linewidth]{FIG/usefullness.png}
%     \caption{Human reranking performance with and with- out category paths on NQ320K dev set pairs where the model retrieves the gold document in the top 5.}
%     \label{fig:usefulness}
% \end{figure}

\subsection{\hype improves retrieval accuracy (RQ1)}
\paragraph{Main results on \nq.}
Table \ref{tab:nq320k} shows retrieval accuracy of various docid types with \hype on \nq.
Overall, \hype consistently improves retrieval accuracy across all docid types in both \textit{seen test} and \textit{unseen test}. 
This demonstrates that \hype's hierarchical category paths can be orthogonally applied to enhance retrieval accuracy across different docid types, suggesting that integrating these paths into existing generative retrieval methods can further improve performance.
While \hype can be applied to all docid types effectively, the experimental results show that \textit{title} docid yields the most significant performance improvement when \hype is applied.
This can be attributed to our analysis that title docids are the most well aligned with hierarchical category paths.
Specifically, hierarchical category paths reflect the semantic structure of document corpus in a coarse-to-fine manner.
Since titles are concise and primarily encode coarse-grained document semantics, they particularly benefit from this hierarchical structure.

% While \hype can be applied to all docid types effectively, the experimental results show that \textit{title} docid yields the most significant performance improvement when \hype is applied. 
% Notably, our hierarchical category paths reflect the semantic structure of document corpus in a coarse-to-fine manner.
% Since titles are concise and primarily encode coarse-grained semantics of document, they particularly benefit from hierarchical structure.
% Since titles are concise and primarily encode coarse-grained document semantics, they particularly benefit from this hierarchical structure.

% Our paths, serve as a pseudo-reasoning, allowing the model to navigate step-by-step through various semantic hierarchical categories before arriving docid. 
% Since titles are concise and inherently reflect a structured overview of a document, they aligns well with the \hype's hierarchical category paths, further enhancing retrieval accuracy. 

\begin{figure}[t]
    \centering
    \includegraphics[width=\columnwidth]{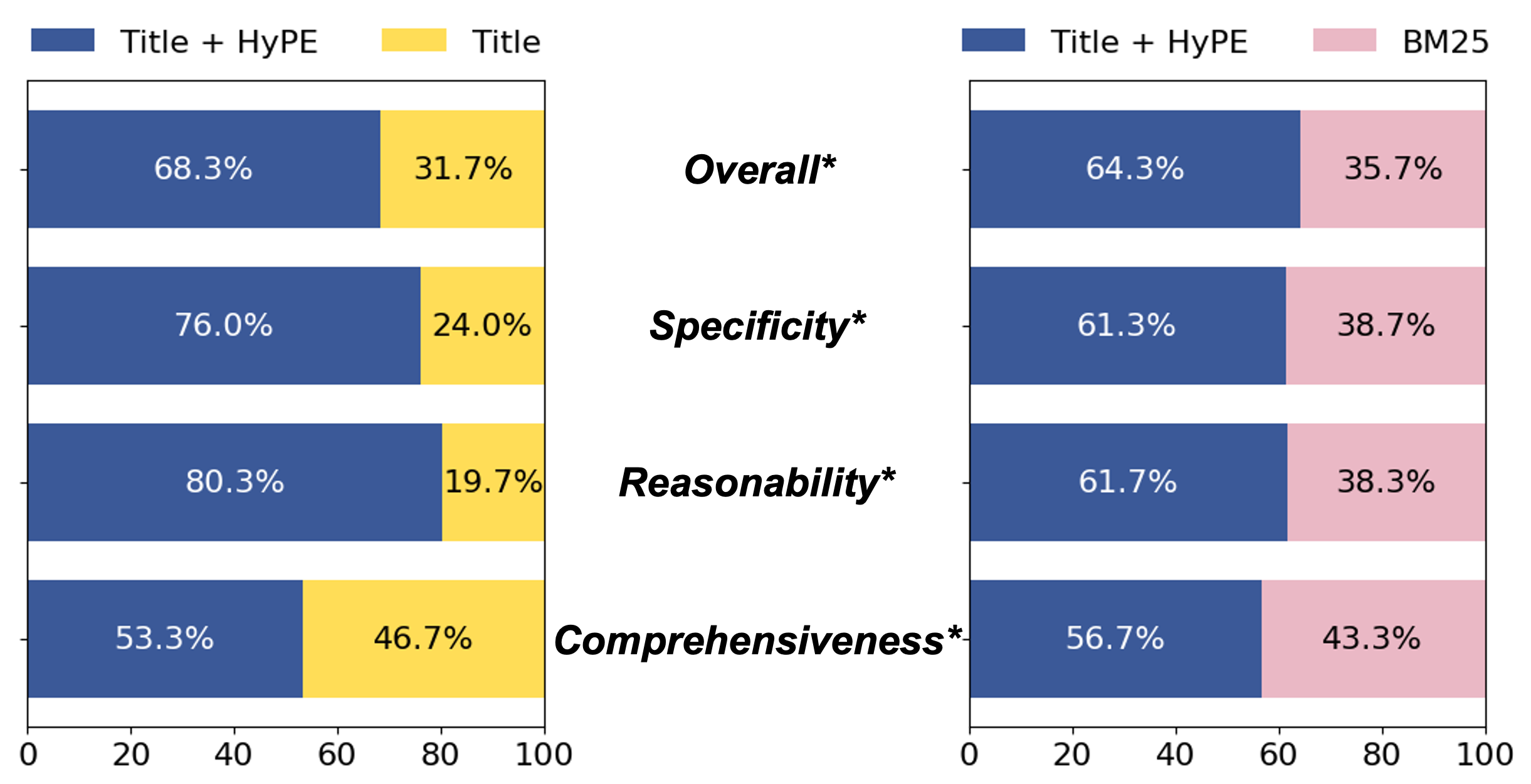}
    \caption{Human evaluation of pairwise comparisons for retrieval explanations between \hype and baselines.}
    \label{fig:humaneval}
\end{figure}

\begin{figure}[t]
    \centering
    \includegraphics[width=\columnwidth]{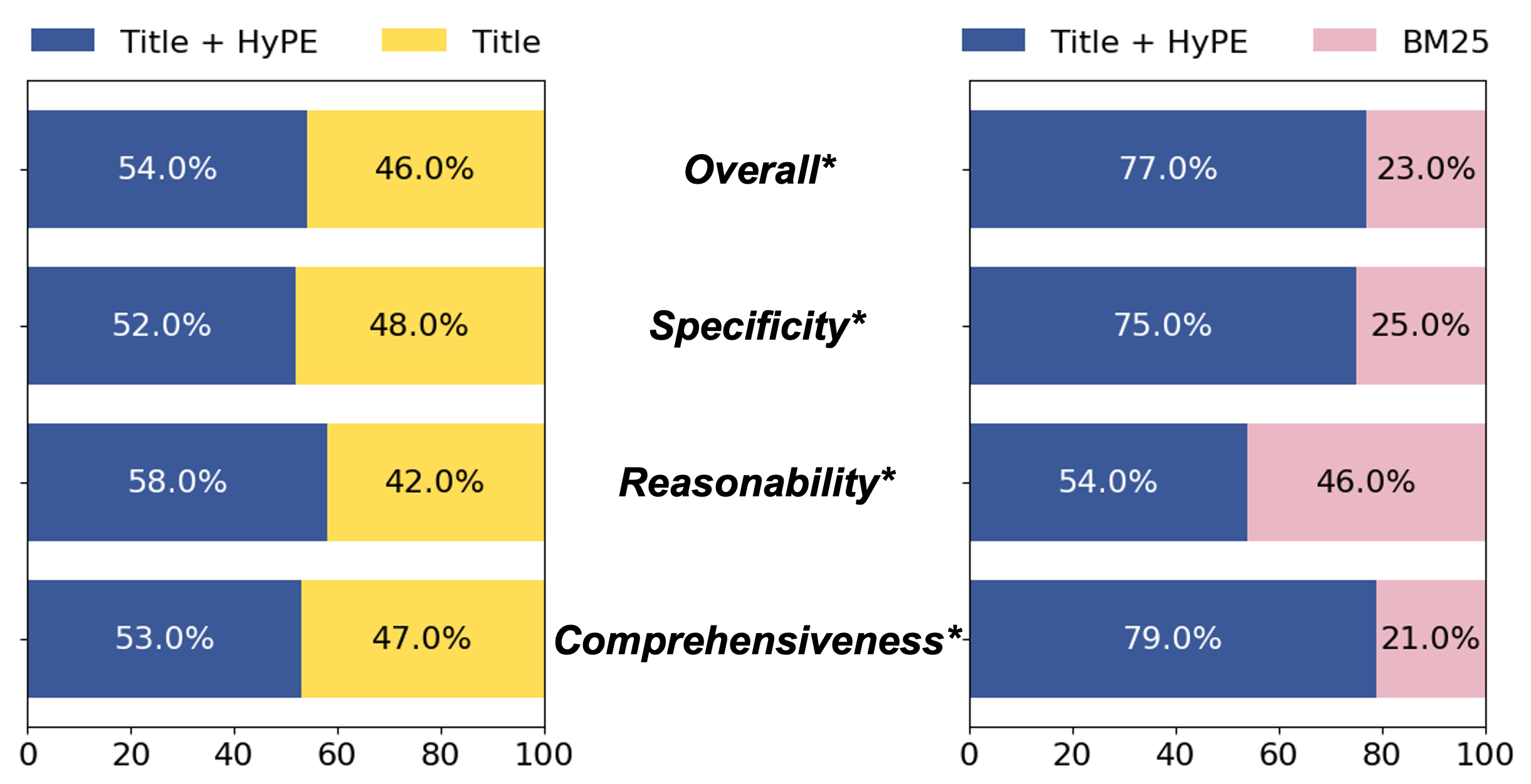}
    \caption{LLM evaluation of pairwise comparisons for retrieval explanations between \hype and baselines.}
    \label{fig:llmneval}
\end{figure}

\paragraph{Results on non-Wikipedia corpus.}
% Since we adopt the Wikipedia category tree as the backbone hierarchy and the \nq is constructed from Wikipedia, the performance gains on \nq could stem from the inherent alignment between the hierarchy and the retrieval corpus.
% To assess this, we further evaluate \hype on \msmarco.
Since both the backbone hierarchy used in this work and the \nq corpus come from Wikipedia, the performance gains on \nq could be influenced by this shared source.
To assess whether these gains generalize beyond Wikipedia-based corpus, we evaluate \hype on \msmarco.
% We therefore evaluate \hype on \msmarco to assess whether its performance gains generalize beyond Wikipedia-based corpus.
% To assess whether the improvements of \hype generalize beyond Wikipedia-based corpus, we evaluate \hype on \msmarco.
Table~\ref{tab:msmarco} shows that \hype consistently improves retrieval accuracy on \msmarco, even when using the same Wikipedia category tree backbone.
% These results suggest that \hype can improve performance by enabling coarse-to-fine pseudo-reasoning guided by a semantic hierarchy without relying on the backbone hierarchy and the retrieval corpus sharing the same source.
These results suggest that \hype improves performance by leveraging coarse-to-fine pseudo-reasoning guided by a semantic hierarchy, with gains that are not limited to cases where the backbone hierarchy and the retrieval corpus have shared source.
% It also suggests the feasibility of plug-and-play replacement of the backbone hierarchy with an in-domain taxonomy when \hype is applied to domain-specific retrieval tasks.
It further demonstrates that \hype can generalize across a broader range of retrieval scenarios.
% This finding also suggests the feasibility of plug-and-play replacement of the backbone hierarchy with an in-domain taxonomy when \hype is applied to domain-specific retrieval tasks.
We leave further discussion of this point to Appendix~\ref{appendix:backbone_category hierarchy}.

\subsection{Hierarchical category paths serve as effective retrieval explanations (RQ2)}
\paragraph{Human evaluation results.}
We evaluate the explanatory quality of the hierarchical category paths of \hype through a human evaluation conducted via Amazon Mechanical Turk (AMT).
% We randomly sample 100 examples from the NQ320K dev set and ask three human judges per sample to compare the quality of the explanations based on four distinct criterias: \textit{overall}, \textit{specificity}, \textit{reasonability} and \textit{comprehensiveness}. 
We ask three human judges per sample to compare the quality of the explanations based on four distinct criterias: \textit{overall}, \textit{specificity}, \textit{reasonability} and \textit{comprehensiveness}. 
For detailed descriptions, please refer to Appendix~\ref{sec:appendixhumaneval}.
In Figure \ref{fig:humaneval}, \hype outperforms both the title docid baseline and BM25 across all criteria.
% , receiving high scores for its \textit{overall} explanation of the retrieval process. 
Specifically, \hype shows a substantial margin of superiority in terms of \textit{specificity} and \textit{reasonability}, demonstrating that \hype provides clearer explanations of retrieval process, as well as more logical and reasonable explanation. 
Furthermore, \hype beats other baselines in \textit{comprehensiveness}, indicating that its hierarchical category path is effective in explaining not only narrow, specific details but also broader semantic information.  

\begin{table}[!t]
\small
\centering

% \resizebox{0.80\linewidth}{!}{
\begin{tabular}{lccc}
\toprule
\textbf{Baseline} & R@1 & M@5 & Conf. \\

% & Query-Output & Document-Output & Overall \\

\midrule
Title Docid
& 19.7 & 47.9 & 4.0 \\

% \quad + \hype (Level 2)
% & 0.49 & 0.51 & 0.49  \\

% \quad + \hype (Level 3)
% & 0.49 & 0.54 & 0.50  \\

\quad + \hype 
& 24.3 & 52.8 & 4.5  \\

\rowcolor{gray!20}
\textbf{Improvement} & 
23.7\% & 10.4\% & 12.0\%  \\

\bottomrule
\end{tabular}
\caption{Human reranking performance with and without category paths on NQ320K dev set pairs where the model retrieves the gold document in the top 5.}
\label{tab:usefulness}
% }
\end{table}

\begin{figure*}[tb!]
    \centering
    \includegraphics[width=\linewidth]{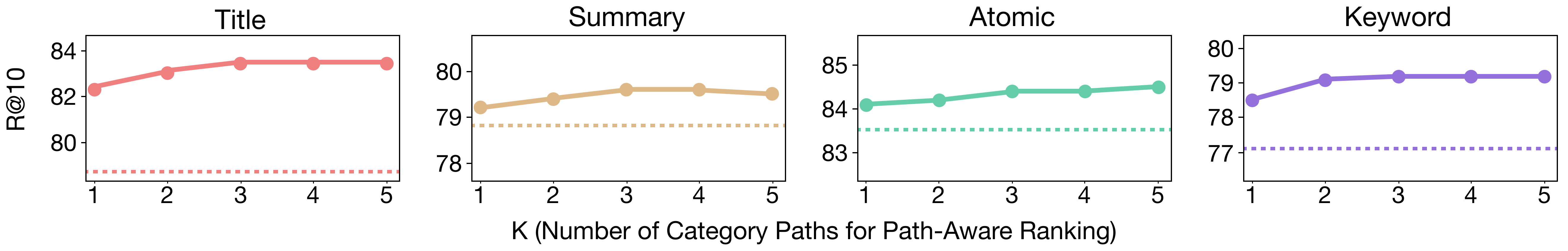}
    % \vspace{-25pt}
    \caption{Performance changes of \hype with respect to the number of generated hierarchical category paths used to construct the ranked docid list. The dashed line indicates the performance of docid-only baseline.}
    \label{fig:path_analysis}
\end{figure*}

% \begin{table*}[t]
%   \centering
%   \small
%   \resizebox{.99\textwidth}{!}{
%   \begin{tabular}{@{} p{0.48\textwidth} p{0.52\textwidth} @{}}
%     \toprule
%     \textbf{Document} & \textbf{Generated Category Paths for Each Query} \\
%     \midrule
%     \begin{minipage}[t]{0.99\linewidth}
%       \textbf{Title}: Sun \\
%       The Sun is the star at the center of the Solar System. \ldots The core is the only region of the Sun that produces an appreciable amount of thermal energy through fusion; \ldots The Sun is about halfway through its main-sequence stage, during which nuclear fusion reactions in its core fuse hydrogen into helium.
%     \end{minipage}
%     &
%     \begin{minipage}[t]{0.99\linewidth}
%       \textbf{Query 1}: the core of the sun in which the sun's thermonuclear energy is produced takes up about \\[0.7ex]
%       \textbf{Generated Category Path}: universe > energy > energy conversion
%       \midrule 
%       \textbf{Query 2}: what stage of the star life cycle is the sun in \\[0.7ex]
%       \textbf{Generated Category Path}: nature > evolution > stellar evolution
%     \end{minipage} \\
%     \bottomrule
%   \end{tabular}
% }
%   \caption{Example of the document annotated for multiple queries in the \nq dev set. The generative retrieval model with \hype generates query-specific category paths based on the topics of the document associated with each query, explaining why the document is retrieved for the particular query.}
%   \label{tab:case_study}
% \end{table*}

\begin{table*}[t]
  \centering
  \small
  \resizebox{.99\textwidth}{!}{
  \begin{tabular}{@{} p{0.48\textwidth} p{0.52\textwidth} @{}}
    \toprule
    \textbf{Document} & \textbf{Generated Category Paths for Each Query} \\
    \midrule
    \begin{minipage}[t]{\linewidth}
      \textbf{Title}: Sun \\
      The Sun is the star at the center of the Solar System. \ldots The core is the only region of the Sun that produces an appreciable amount of thermal energy through fusion; \ldots The Sun is about halfway through its main-sequence stage, during \ldots
    \end{minipage}
    &
    \begin{minipage}[t]{\linewidth}
      \textbf{Q1}: core of the sun in which the sun’s thermonuclear energy \ldots \\[0.7ex]
      \textbf{Generated Category Path}: universe > energy > energy conversion \\[-1.0ex]
      \rule{\linewidth}{0.4pt}
      \textbf{Q2}: what stage of the star life cycle is the sun in \\[0.7ex]
      \textbf{Generated Category Path}: nature > evolution > stellar evolution
    \end{minipage}
    \\
    \bottomrule
  \end{tabular}
  }
  \caption{Example of the generated query-specific hierarchical category paths for the same document from \nq dev set. \hype generates explanations based on the topics of the document associated with each query.}
  \label{tab:case_study}
\end{table*}

\paragraph{LLM evaluation results.}
While human evaluation is the gold standard for evaluating explanation quality, recent works~\cite{kim-etal-2024-prometheus, tan2025judgebench} have shown that LLM-based evaluation can serve as a scalable and reasonably well-aligned proxy for human judgment. 
Motivated by this, we additionally conduct an LLM-based evaluation using GPT-5 as the judge. 
Specifically, We replicate the exact same pairwise comparison setup used in the human evaluation, providing the LLM judge with identical instructions and criteria used by human judges. 
For more detailed descriptions, please refer to Appendix~\ref{sec:appendixhumaneval}.
As shown in Figure~\ref{fig:llmneval}, \hype is consistently preferred over both the title docid baseline and BM25 across all four criteria. 
Importantly, these results closely align with the human evaluation results, confirming that the explainability of hierarchical category paths is robust across evaluation methods.
Overall, these findings demonstrate that hierarchical category paths can serve as effective retrieval explanations.

% Overall, these findings clearly imply that our explanations which utilize hierarchical category paths significantly improve the clarity and
% usefulness of retrieved results.
% \textbf{These results highlight that \hype's pseudo-reasoning, which utilizes hierarchical category paths, provides users with a effective explanation of the retrieval process.} 

\subsection{\hype guides users in making better search decision by explanations (RQ3)}
% In real-world search systems, users are typically provided only with the document title and the first few lines when deciding which result to open. 
We investigate whether explanations of \hype can help users effectively identify relevant documents in realistic search settings, where decisions are often based only on titles and short snippets~\cite{10.1145/1076034.1076063}.
To this end, we conduct a human reranking experiment via AMT using the NQ320K dev set. 
Specifically, human judges rerank the top-5 retrieved results by relevance and rate their confidence (1–5) under two settings: title only, and title with category path.
% With human-reranking results, we measure performance with Recall@1, MRR@5 and \textit{Confidence}.
We measure performance with Recall@1, MRR@5 and \textit{Confidence}.
More details are provided in Appendix~\ref{sec:appendixhumanrerank}.

Table~\ref{tab:usefulness} shows that providing category paths as explanations leads to substantial improvements in human reranking accuracy, with Recall@1 and MRR@5 increasing by 23.7\% and 10.4\%, respectively.
This suggests that explanations of \hype help users better understand retrieval results and select relevant documents.
Furthermore, the availability of category paths increases user confidence by 12.0\%.
Together, these findings indicate that \hype explanations provide both accurate selections and decision confidence in realistic search scenarios, where relevance judgments are often made based on limited signals such as titles or short snippets.

\section{Analysis}
\label{sec:analysis}
% \subsection{Case Study}

% \subsection{Analysis of Path-Aware Reranking}
\paragraph{Analysis of category path effectiveness.}
As shown in Figure~\ref{fig:path_analysis}, we analyze the effectiveness of incorporating category paths in \hype by varying the number of paths under the \textit{\pars} strategy and evaluating retrieval performance.
Overall, using category paths consistently outperforms the docid-only setting regardless of $K$. 
Even the single-path setting ($K=1$) improves performance compared to using docids alone. 
% (see Appendix~\ref{appendix:ablationforpaths} for full results).
% To validate the effectiveness of \textit{\pars} strategy, we analyze the performance changes in retrieval accuracy with respect to the number of hierarchical category paths considered by \hype. 
% Figure~\ref{fig:path_analysis} presents the analysis results. 
% First, \textbf{even the single-path setting ($K=1$), which does not use \pars, already improves retrieval accuracy compared to using docids alone} (see Appendix~\ref{appendix:ablationforpaths} for full results). 

\paragraph{Analysis of path-aware ranking.}
Moreover, incorporating additional paths further boosts performance across all baselines, revealing a clear gap between the without-\pars ($K=1$) and with \pars ($K>2$).
These results indicate that considering multiple paths through the \pars strategy allows the most relevant docids to be prioritized in the final ranked list, thereby enhancing retrieval accuracy.
However, we observe that using too many paths eventually leads to a plateau in performance improvement. 
Beyond a certain threshold, additional paths tend to introduce noise or increase unnecessary complexity. 
% Consequently, using three paths achieves optimal retrieval accuracy for most docid types.

\paragraph{Case study.}
% We conduct two types of case studies to verify our \hype provides more effective explanations for each query.
% Table~\ref{tab:case_study} illustrates \hype's explanations in cases where a single document is annotated with multiple queries on different topics. 
Table~\ref{tab:case_study} illustrates examples of explanations generated by \hype for different queries targeting different topics within the same document.
For the query ``\textit{the core of the sun in which the sun's thermonuclear energy is produced}'', \hype generates paths related to the universe and energy conversion, clearly explaining the thematic relevance between the query and the document. 
However, for another query, ``\textit{what stage of the star life cycle is the sun in}'', it generates a path related to stellar evolution, which is different from the previously generated path but relevant to the query. 
This shows that \hype can provide effective explanations to users  by tailoring them to each query.

\paragraph{Analysis of error propagation.}
Since \hype generates hierarchical category paths prior to docid decoding, incorrectly generated paths could potentially cause error propagation, leading to the decoding of irrelevant docids. 
To examine whether \hype actually suffers from such error propagation, we evaluate the quality of generated paths and analyze their impact on retrieval accuracy. 
Specifically, for queries where the title docid baseline successfully retrieves the gold document within its top-100 ranked list, we run \hype on these queries to generate category paths and decode docids. 
We then employ GPT-5 as an evaluator to make binary judgments on whether the generated path appropriately explains the relationship between the query and the decoded docid.
Based on this judgment, we split the queries into two groups, Relevant path and Irrelevant path, and compare their R@1 and R@10 performance.
Table~\ref{tab:error_propagation} shows that 94.6\% of generated paths are judged as relevant, indicating that \hype produces meaningful explanations for the vast majority of queries. 
For the relevant path group, \hype outperforms the baseline on both R@1 and R@10, confirming that appropriately generated paths make a substantive contribution to retrieval quality. 
Notably, even in the irrelevant path group, performance does not drop significantly, and R@10 rather increases.
These findings demonstrate that \hype is robust to path generation errors without severe error propagation.

\begin{table}[t]
\centering
\small
\setlength{\tabcolsep}{4pt}
\begin{tabular}{cccccc}
\toprule
\multirow{2.5}{*}{\shortstack{Path\\Relevance}} & \multirow{2}{*}{N} & \multicolumn{2}{c}{R@1} & \multicolumn{2}{c}{R@10} \\
\cmidrule(lr){3-4} \cmidrule(lr){5-6}
 & & Baseline & +HyPE & Baseline & +HyPE \\
\midrule
Relevant & 6,514 & 70.4 & \textbf{71.5} & 89.5 & \textbf{92.7} \\
Irrelevant & 372 & \textbf{75.0} & 73.9 & 89.8 & \textbf{90.1} \\
\bottomrule
\end{tabular}
\caption{Retrieval performance comparison between relevant and irrelevant path groups on NQ320K dev set, using title docid. Path relevance is judged by GPT-5.}
\label{tab:error_propagation}
\end{table}

% \begin{table}[t]
% \centering
% \small
% \begin{tabular}{lccc}
% \toprule
%  & Relevant & Irrelevant \\
% \midrule
% N & 6,514 (94.6\%) & 372 (5.4\%) \\
% \midrule
% Baseline R@1 & 70.4\% & 75.0\% \\
% HyPE R@1 & 71.5\% & 73.9\% \\
% % \midrule
% Baseline R@10 & 89.5\% & 89.8\% \\
% HyPE R@10 & 92.7\% & 90.1\% \\
% \bottomrule
% \end{tabular}
% \caption{Retrieval performance comparison between relevant and irrelevant path groups on NQ320K, using title docid. Path relevance is judged by GPT-5.}
% \label{tab:error_propagation}
% \end{table}
% \subsection{Analysis of Efficiency}
\paragraph{Analysis of inference-time efficiency.}
Providing explanations in \gr inevitably introduces additional computation beyond decoding docids alone, as the generative retrieval model must generate explanatory content in addition to the docid~\cite{jia2025improving}. 
As a result, explainable retrieval methods should provide informative explanations while maintaining minimal computational overhead.
To assess whether \hype satisfies this requirement, we measure the average per-instance inference time under two settings: decoding only docids and decoding docids with path generation stage. 
The full setup is described in Appendix~\ref{appendix:efficientanalysis}.
As shown in Table~\ref{tab:efficiency}, incorporating path generation results in only a modest increase in inference time compared to decoding docids alone. 
This result indicates that \hype require minimal overhead yet enhance both explainability and performance.

\begin{table}[!t]
\centering
\small
\begin{tabular}{cccc}
\toprule
\textbf{Docid Type} & \textbf{Docid Only} & \textbf{Docid + \hype} \\
\midrule
Summary & 0.8127s & 0.9134s \\
Keyword & 1.0389s & 1.1402s  \\
\bottomrule
\end{tabular}
\caption{Average inference time per instance for decoding only docid vs decoding both docid and a single path.}
\label{tab:efficiency}
\end{table}
\begin{table}[!t]
\centering
\small
\begin{tabular}{lcc}
\toprule
\textbf{Explanation Format} & \textbf{Avg. \#Tokens} & \textbf{Efficiency} \\
\midrule
Natural Language & 61.0 & - \\
Category Path (ours) & \textbf{13.5} & \textbf{4.5$\times$} \\
\bottomrule
\end{tabular}
\caption{Average token count of explanations. \hype's hierarchical category paths are approximately 4.5$\times$ more efficient than natural language explanations.}
\label{tab:efficiencytoken}
\end{table}
\paragraph{Analysis of token-level efficiency.}
Furthermore, we compare the token-level efficiency of \hype's category paths against natural language explanations. 
We randomly sample 300 query-document pairs from the training set and generate natural language relevance explanations using GPT-5. 
As shown in Table~\ref{tab:efficiencytoken}, \hype's category paths average only 13.5 tokens, while natural language explanations average 61.0 tokens, making category paths approximately 4.5$\times$ more efficient. 
This compact structure also reduces inference latency, further highlighting the practicality of \hype.
% This indicates that explanations based on our hierarchical category paths require only minimal additional computational overhead. 
% Even with this minimal overhead, these explanations effectively enhance both explainability and retrieval performance, as demonstrated in the preceding sections.8

% However, the minor increase in inference time is justified by the explainability and overall performance gains that HyPE offers.
% Our experimental results show that the path generation stage has a negligible impact on efficiency, while HyPE consistently improves retrieval accuracy and explainability across various settings.

% \section{Related Work}
% \label{sec:relwork}
% \input{020relatedwork}

\section{Conclusion}
\label{sec:conclusion}
%\underline{\textbf{H}}ierarchical Categor\underline{\textbf{y}} \underline{\textbf{P}}ath \underline{\textbf{E}}nhanced Generative Retrieval. 
% By leveraging hierarchical category paths, \hype addresses the limitations of existing \gr methods that struggle with providing meaningful explanations for why specific documents are retrieved.
% Furthermore, we introduce \textit{\pars} allows for the aggregation of diverse topic information, ensuring that the most relevant docids are prioritized in the final ranked list of docids.
% Our experiments demonstrate that \hype not only enhances overall retrieval performance but also helps users make more accurate decisions during search by providing effective explanations.

In this paper, we propose \hype, a framework designed to enhance the explainability of generative document retrieval by utilizing hierarchical category paths. 
Our experiments demonstrate that the hierarchical category paths used in \hype can serve as effective retrieval explanations, which not only enhance overall retrieval performance but also help users make more accurate decisions during realistic search scenarios.
Moreover, these benefits are achieved with minimal additional computational overhead, offering a practical approach for explainable retrieval system.
% To the best of our knowledge, this work is the first attempt to enhance explainability in \gr.
We hope our work advances the development of explainable retrieval systems.
% We hope our work advances the development of explainable retrieval systems that improve both reliability and practical usability.
% To the best of our knowledge, this work is the first attempt to enhance explainability in \gr, and we hope our research paves the way for positive advancements in this field.
% To the best of our knowledge, this work is the first attempt to enhance explainability in \gr, and we hope our research paves the way for positive advancements in this field.
% We hope our research paves the way for positive advancements in this field.
% We hope our research paves the way for meaningful progress in the development of retrieval systems.

% \hype prepends paths which extracted from high-quality and interpretable semantic structure to the front of the docid, effectively reflecting the semantic structure without any semantic impurity.

\section*{Limitations}
Despite the promising results and contributions of \hype, our work has four key limitations stemming from computational costs and budget constraints. 
First, we do not experiment with alternative backbone hierarchies beyond Wikipedia’s category tree. 
While it is possible that domain-specific taxonomies may further improve retrieval performance in specialized settings, we consider Wikipedia’s broad and deep hierarchy sufficient for general-purpose document retrieval. 
Please refer to Appendix~\ref{appendix:backbone_category hierarchy} for further discussion.
Second, due to cost and scalability constraints, we do not conduct human evaluations to assess how different path depths affect the quality of the explanation.
Instead, we provide a limited analysis of explainability with respect to path depth using STS score in the Appendix~\ref{appendix:sts}. 
Third, we evaluate \hype using a basic generative retrieval setup (Section \ref{sec:preliminary}) to isolate its effect. 
We do not incorporate advanced optimization techniques or architectures from recent works, which may further improve performance of \hype.
Fourth, we evaluate \hype using T5-base as the backbone, following the convention in generative retrieval~\cite{nips/Tay/DSI, corr/abs-2304-04171/GENRET} or recommendation~\cite{kim2024mviger, lee-etal-2025-gram} works. 
Since generative retrieval requires the model to memorize the entire document corpus and perform constrained beam search during evaluation, directly applying billion-scale models such as Llama remains challenging under conventional settings. 
For this reason, T5-base remains the de facto backbone even in recent generative retrieval works~\cite{zhang-etal-2025-multi-level, 10.1145/3726302.3730314, 10.1145/3726302.3730023}. While we believe \hype can generalize to other backbones, we leave verification on larger models as future work.

\section*{Ethical Statement}
This study strictly adhered to ethical guidelines throughout the human evaluation and data usage process. 
All content used in the human evaluation and human reranking—including NQ320K and Wikipedia documents—was publicly accessible and did not involve any private or proprietary data. 
We did not obtain IRB approval for our study, following precedents set by prior work~\cite{kim-etal-2023-soda, kang-etal-2024-large} which conducted similar human evaluations without IRB oversight. 
We ensure that no ethical concerns would arise during the evaluation.
The evaluation and reranking were conducted on Amazon Mechanical Turk (AMT), where all participation was anonymous and no personal information was collected at any stage.
For human evaluation, we hire three different judges per instance from AMT and guarantee fair compensation for each judge. We pay \$0.15 for each unit task.
Human judges were fully informed about the task's purpose, procedure, and estimated time requirement before beginning the task. 
All examples were screened to exclude offensive, hateful, or sensitive content and were limited to socially and culturally neutral topics.
All datasets are publicly available and appropriately licensed; the NQ dataset under the Apache 2.0 license and the MS MARCO dataset under the MIT license.

\section*{Acknowledgments}
This work was supported by the IITP grants funded by the Korea government (MSIT) (No. RS-2020-II201361; RS-2024-00457882, AI Research Hub Project; IITP-2026-RS-2020-II201819).

% This document has been adapted
% by Steven Bethard, Ryan Cotterell and Rui Yan
% from the instructions for earlier ACL and NAACL proceedings, including those for
% ACL 2019 by Douwe Kiela and Ivan Vuli\'{c},
% NAACL 2019 by Stephanie Lukin and Alla Roskovskaya,
% ACL 2018 by Shay Cohen, Kevin Gimpel, and Wei Lu,
% NAACL 2018 by Margaret Mitchell and Stephanie Lukin,
% Bib\TeX{} suggestions for (NA)ACL 2017/2018 from Jason Eisner,
% ACL 2017 by Dan Gildea and Min-Yen Kan,
% NAACL 2017 by Margaret Mitchell,
% ACL 2012 by Maggie Li and Michael White,
% ACL 2010 by Jing-Shin Chang and Philipp Koehn,
% ACL 2008 by Johanna D. Moore, Simone Teufel, James Allan, and Sadaoki Furui,
% ACL 2005 by Hwee Tou Ng and Kemal Oflazer,
% ACL 2002 by Eugene Charniak and Dekang Lin,
% and earlier ACL and EACL formats written by several people, including
% John Chen, Henry S. Thompson and Donald Walker.
% Additional elements were taken from the formatting instructions of the \emph{International Joint Conference on Artificial Intelligence} and the \emph{Conference on Computer Vision and Pattern Recognition}.


% \balance
\bibliography{reference}

\clearpage
\newpage
% \newpage
\appendix
% \section{Appendix}
\section{Appendix}
\subsection{Quantitative Analysis of Explainability}
\label{appendix:sts}
We quantitatively evaluate whether \hype's hierarchical category path provides a valid explanation by effectively capturing the semantic relationship between the query and the document. 
To this end, we use a semantic textual similarity (STS) model~\cite{agirre-etal-2012-semeval}\footnote{We use sentence-transformers/roberta-base-nli-stsb-mean-tokens as STS model.} to measure the semantic relevance between two sentences, evaluating the semantic relevance between the query and explanation, as well as between the document and explanation.
Specifically, for each baseline, we use the model output as an explanation and calculate the STS scores for both the query-explanation and document-explanation pairs. 
We then compute the geometric mean of these two scores to evaluate how effectively the explanation captures the relationship between the query and the document.
To further analyze the role of hierarchical category paths in explainability, we consider how varying the maximum level of the paths impacts semantic relevance.
As mentioned in Section \ref{subsubsec:pathcollection}, \hype basically leverages Level 4 paths, but we also experiment with varying the maximum level (e.g., Level 2, Level 3) to examine how the maximum level of paths influences the explainability of the query-document relationship. 
In addition, we also include BM25 as a baseline, which is capable of providing explanations for its retrieval results.  
For the explanation of BM25, we consider the top-3 terms that have the highest BM25 scores calculated between a given query and a document.
\begin{table}[h]
\small
\centering
% \resizebox{0.80\linewidth}{!}{
\begin{tabular}{lcccc}
\toprule
\multirow{2.5}{*}{\textbf{Baseline}} 
& \multicolumn{3}{c}{\textbf{Semantic Relevance}} \\

\cmidrule(lr){2-5}
% & Query-Output & Document-Output & Overall \\
& Query & Document & Overall \\

\midrule
Title Docid
& \underline{0.52} & 0.46 & 0.48 \\

\quad + \hype (Level 2)
& 0.49 & 0.51 & 0.49  \\

\quad + \hype (Level 3)
& 0.49 & 0.54 & 0.50  \\

\quad + \hype 
& 0.50 & \underline{0.56} & \underline{0.52}  \\
\midrule

Keyword Docid
& 0.42 & 0.54 & 0.47   \\

\quad + \hype (Level 2)
& 0.41 & 0.56 & 0.47  \\

\quad + \hype (Level 3)
& 0.41 & 0.57 & 0.47  \\

\quad + \hype 
& \underline{0.43} & \underline{0.58} & \underline{0.49} \\

\midrule

Summary Docid
& \underline{0.46} & 0.69 & 0.55   \\

\quad + \hype (Level 2)
& 0.45 & 0.70 & 0.55  \\

\quad + \hype (Level 3)
& 0.45 & 0.70 & 0.55  \\

\quad + \hype 
& 0.45 & \underline{\textbf{0.71}} & \underline{\textbf{0.57}} \\

\midrule

BM25
& \textbf{0.56} & 0.31 & 0.42  \\

\bottomrule
\end{tabular}
% }
\caption{Semantic relevance between query/explanation and document/explanation on 1,000 \nq dev set pairs where each baseline successfully retrieves the relevant document at rank 1.}
\label{tab:sts}
\end{table}

As shown in Table \ref{tab:sts}, applying \hype improves overall semantic relevance across all baselines. 
This indicates that \hype's category path effectively captures and explains the relationship between the query and the document. 
We note that \hype achieves higher overall relevance than the term-matching method (i.e., BM25), further proving the validity of the \hype's category path as an explanation. 
Moreover, maximum level of hierarchical category path significantly influences overall semantic relevance. 
Specifically, paths with fewer levels than the default level (level 4) fail to capture sufficient semantic relevance between the query and the document, resulting in limited explainability. 
These results demonstrate that for category paths to effectively serve as explanations, they must achieve \textit{specificity} necessary to sufficiently explain specific and detailed semantic information, as mentioned in Section \ref{subsec:stepone}.

\subsection{Pseudo-Reasoning}
\label{appendix:pseudo_reasoning}
Generating the hierarchical path resembles step-by-step reasoning. 
However, unlike natural language-based reasoning in \llm, we use the term ``pseudo-reasoning'' because the path structure is more akin to pseudo-code.

\subsection{Backbone category hierarchy}
\label{appendix:backbone_category hierarchy}

\paragraph{Criteria for Selecting the backbone.}
To address the criteria mentioned in Section \ref{subsec:stepone}—Semantic Hierarchy, Generalizability, and Specificity—we utilize Wikipedia’s category tree as the foundation for our hierarchical structure, designating the Main Topic classification category as the root node of the hierarchy. 

\begin{itemize}[leftmargin=*,topsep=2pt,itemsep=2pt,parsep=0pt]
    \item \textit{Semantic Hierarchy}: Are they semantically hierarchical, allowing step-by-step progression in the generation process to clearly represent a specific semantic level?
    \item \textit{Generalizability}: Are they able to provide semantic information across a wide range of domains? 
    \item \textit{Specificity}: Are they capable of sufficiently explaining specific and detailed information? 
\end{itemize}

% \begin{table}[t!]
%     \centering
%         \label{tab:hierarchystatistics}
%     \begin{tabular}{cccccc}
%     \toprule
%     \textbf{} & \textbf{Level 1} & \textbf{Level 2} & \textbf{Level 3} & \textbf{Level 4} & \textbf{Total} 
%     \\
%     \midrule
%     {\# Nodes} & 40 & 1,330 & 13,383 & 95,240 & 109,993 \\
%     \bottomrule
%     \end{tabular}
%     \caption{Statistics of the hierarchical category.}
% % \vspace{-30pt}
% \end{table}

\begin{table}[h]
    \centering
    \small
    \label{tab:hierarchystatistics}
    \begin{tabular}{ccccc}
    \toprule
    \textbf{Level 1} & \textbf{Level 2} & \textbf{Level 3} & \textbf{Level 4} & \textbf{Total} 
    \\
    \midrule
    40 & 1,330 & 13,383 & 95,240 & 109,993 \\
    \bottomrule
    \end{tabular}
    \caption{Statistics of the used category hierarchy, showing the number of nodes at each level (or depth).}
% \vspace{-30pt}
\end{table}

\paragraph{Wikipedia category tree overview.}
Wikipedia’s category tree consists of 40 nodes at level 1, covering broad categories such as \textit{Business, Sports, Science, Philosophy, Language, Health, Government, Culture}, and others. This feature of encompassing a wide range of fields ensures that Wikipedia's category tree satisfies the criterion of \textit{Generalizability}, as it can be applied across various domains. 
Moreover, these broad categories are further subdivided into increasingly specific subcategories as the level increases. 
For instance, level 1 \textit{Science} is divided into major subcategories such as \textit{Branches of Science, Scientists}, and \textit{History of Science} at level 2. 
Among these, \textit{Branches of Science} is further refined into \textit{Applied Science, Formal Science}, and \textit{Social Science} at level 3, which are then expanded into even more specific subcategories like \textit{Computer Science, Agronomy, Metrology}, and \textit{Bioinformatics} at level 4. 
As the levels progress, the structure captures increasingly detailed semantic information, effectively fulfilling the criterion of \textit{Specificity}. 
Additionally, the broad-to-specific hierarchical structure of Wikipedia's category tree naturally achieves \textit{Semantic Hierarchy}.

\paragraph{Implementation details for path.}
To utilize Wikipedia’s category tree, we employed Selenium\footnote{https://pypi.org/project/selenium/} to recursively scrape the Wikipedia and extract the Wikipedia category tree. 
When linearizing the category hierarchy into a hierarchical category path, each category is connected using the delimiter >. 
The delimiter > is chosen among several candidate delimiters because it showed the highest semantic similarity to the natural language sentence ``\textit{the right category is included in the left category}'', as measured by Sentence-T5. 

\paragraph{Scalability of our backbone hierarchy.}
We believe that Wikipedia’s category tree will function effectively in most document retrieval scenarios. This taxonomy was specifically designed to systematically categorize real Wikipedia documents, which cover a wide range of domains and knowledge. 
\textbf{Its broad and deep structure ensures that it can encompass diverse domains effectively, making it a strong backbone hierarchy for general-purpose retrieval systems.} 

\paragraph{Adaptability of \hype.}
However, we acknowledge that in more specialized domains—such as expert-driven fields like medicine, law, or scientific literature—the Wikipedia-based hierarchy may not fully capture domain-specific semantics or categorization needs. 
In such cases, the backbone hierarchy may need to be replaced or augmented with a domain-specific taxonomy better suited to the task. 
\textbf{We note that \hype is compatible with this setting: domain-specific taxonomies can be integrated in a plug-and-play fashion. }
For example, the domain taxonomy used for academic paper retrieval~\cite{Kang2024TaxonomyguidedSI} could be adopted as an alternative backbone in that context.
Furthermore, if a well-defined taxonomy does not yet exist for a specific domain, one can be constructed using taxonomy induction methods~\cite{Zhang2018TaxoGenUT, Lee2022TaxoComTT}.

\subsection{Details of the path augmented training set}
\label{appendix:detailtrainingset}
We filter out path set for each document $D$ by leveraging Sentence T5~\cite{acl/NiACMHCY22/SentenceT5} as bi-encoder.
Also, we leverage LLM\footnote{We use Llama-3-8B-Instruct~\cite{Dubey2024TheL3} as \llm.} to generate the final path set $\mathcal{P}_{D}$, selecting up to three paths that best represent the document.

\subsection{Dataset Overview}
\label{appendix:dataset_overview}
\begin{table}[t]
\centering
\small
\resizebox{.99\columnwidth}{!}{
\begin{tabular}{cccc}
\toprule
\textbf{Dataset} & \textbf{\# Docs} & \textbf{\# Train queries} & \textbf{\# Test queries} \\
\midrule
\nq & 109,739 & 307,373 & 7,830 \\
\msmarco & 323,569 & 366,235 & 5,187 \\
\bottomrule
\end{tabular}
}
\caption{Statistics of the document retrieval datasets used.}
\label{tab:datasetOverview}
\end{table}
In this work, we use \nq and \msmarco.
For \nq, we follow NCI~\cite{nips/WangHWMWCXCZL0022/NCI} setup and adhered to the seen and unseen test splits used in GENRET~\cite{corr/abs-2304-04171/GENRET}. 
For \msmarco, we construct dataset based on the MSMARCO document ranking dataset, following setups from Ultron~\cite{corr/abs-2208-09257/Ultron}, GENRET~\cite{corr/abs-2304-04171/GENRET}, and NOVO~\cite{CIKM/Wang2023/NOVO}. 
% Methods using the MSMARCO passage ranking dataset, such as TOME~\cite{acl/RenLWWW23/TOME} and GLEN~\cite{Lee2023GLENGR}, or using NQ100K like SE-DSI~\cite{SIGKDD/Tang2023/SE-DSI}, are excluded from consideration as reference baselines unless they report results specifically on the our dataset setup. 
Table \ref{tab:datasetOverview} shows the statistical details of the datasets used in our experiments. 

\subsection{Human Evaluation}
\label{sec:appendixhumaneval}
We assess the quality of the generated explanations by conducting a human evaluation, where we compare the outputs of \hype to other baseline models using Amazon Mechanical Turk (AMT). 
In this experiment, we use the title docid baseline described in Section~\ref{subsec:experimentsetup}, and additionally include BM25 as a baseline. which is capable of providing explanations for its retrieval results by highlighting the top-ranked terms contributing to the retrieval.
We ask human judges to evaluate each sample's explanations based on the following four criteria.

\begin{figure*}[t]
    \centering
    \includegraphics[width=\textwidth]{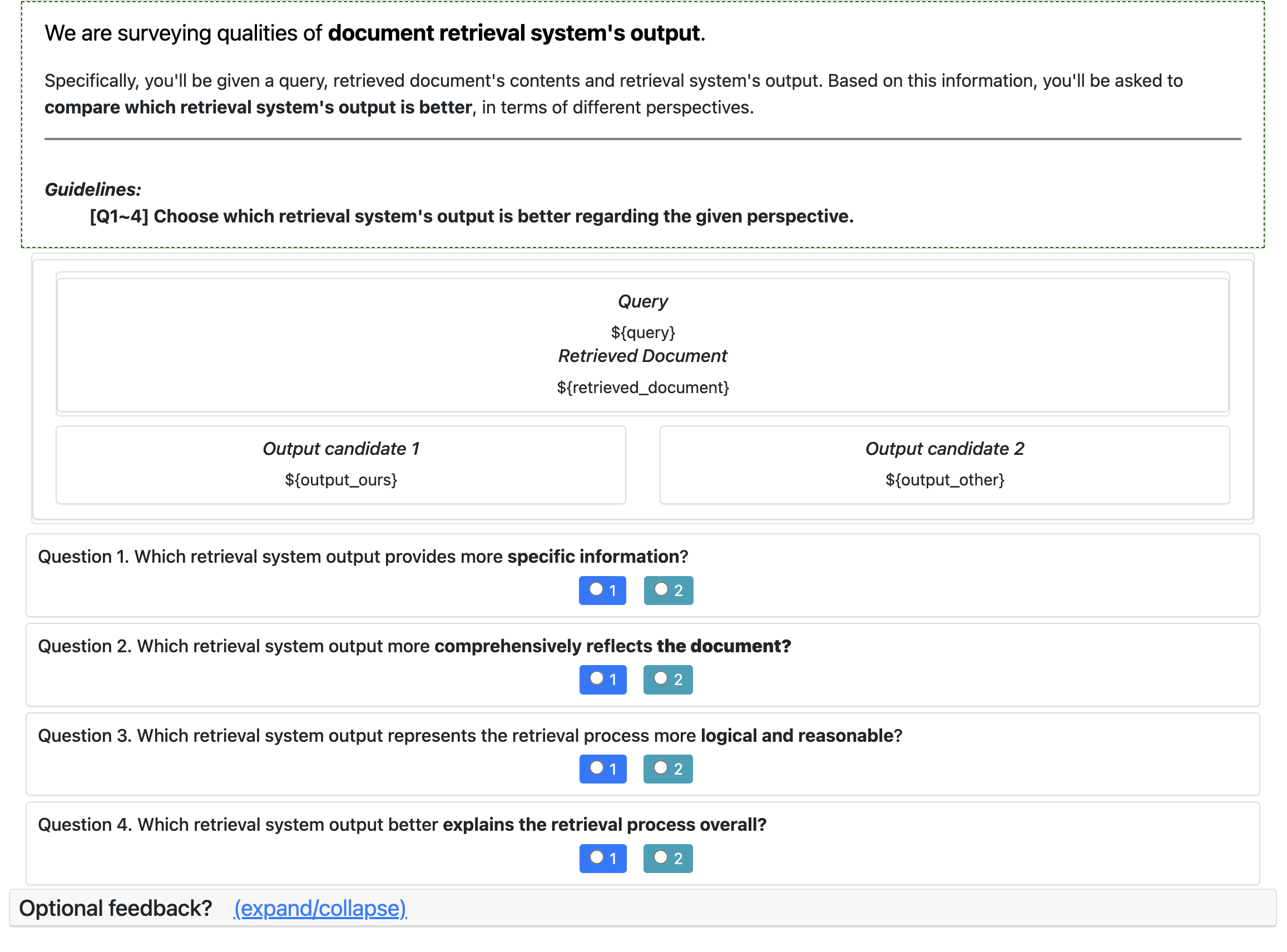}
    \caption{Annotator interface of human evaluation on retrieval system output.}
    \label{fig:humanevalinterface}
\end{figure*}

\begin{itemize}[leftmargin=*,topsep=2pt,itemsep=2pt,parsep=0pt]
    \item 
    \textbf{Overall}: Which retrieval system output better explains the retrieval process overall?

    \item 
    \textbf{Specificity}: Which retrieval system output provides more specific information?

    \item 
    \textbf{Reasonability}: Which retrieval system output represents the retrieval process more logically and reasonably?

    \item 
    \textbf{Comprehensiveness}: Which retrieval system output more comprehensively reflects the content of the document?

\end{itemize}
Note that our human evaluation involved a total of 300 human judges, with each sample being independently evaluated by 3 different human judges. 
This setting is designed by referencing previous works that conduct human evaluation~\cite{kim-etal-2023-soda, Lee2025ImagineAT, kim-etal-2025-towards}.
We show the interface for the human evaluation in Figure~\ref{fig:humanevalinterface}

\subsection{Human Reranking}
\label{sec:appendixhumanrerank}
\begin{figure*}[t]
    \centering
    \includegraphics[width=0.9\textwidth]{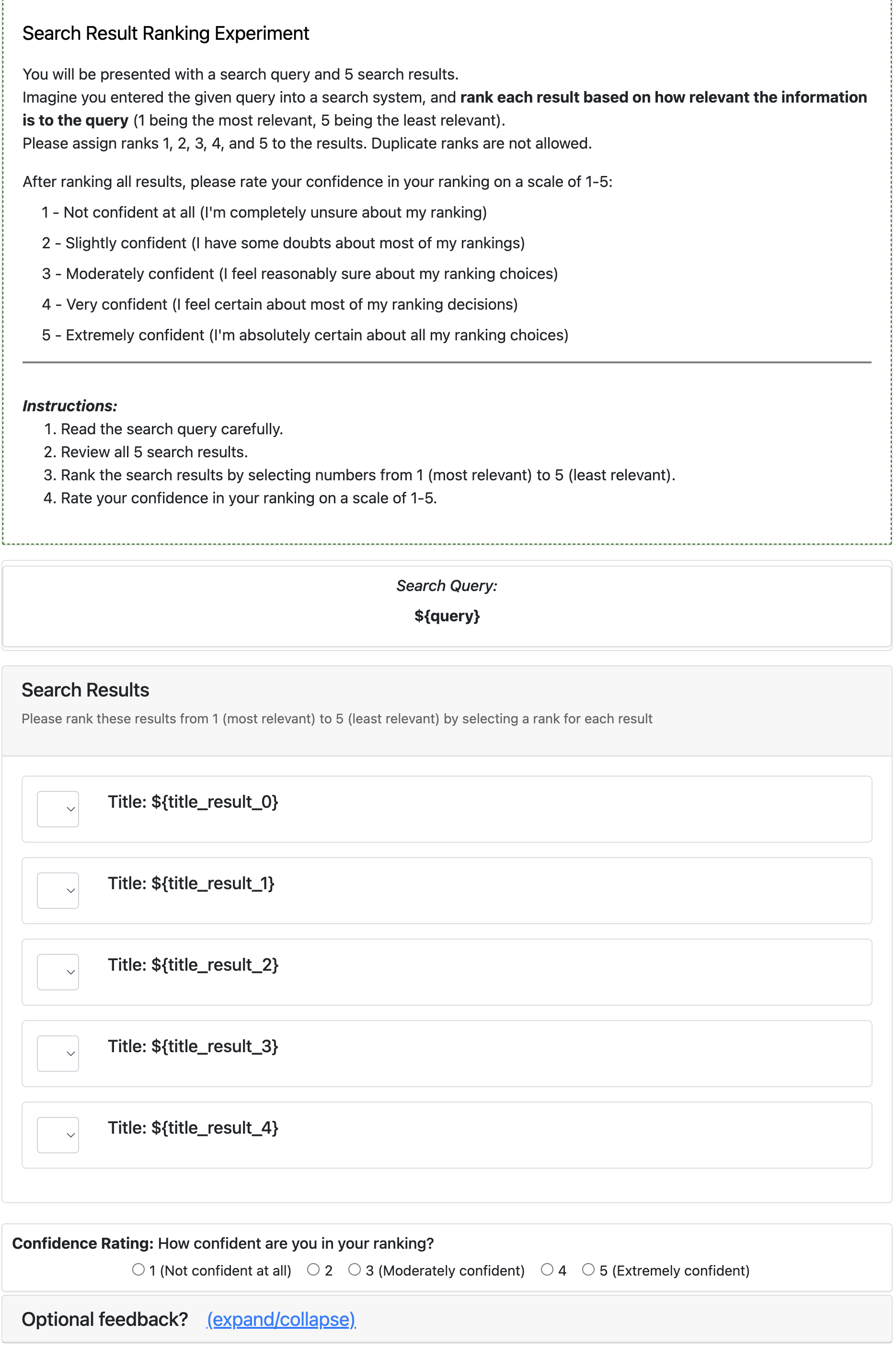}
    \caption{Annotator interface of human reranking on retrieval system output.}
    \label{fig:humanrerankinginterface}
\end{figure*}

To evaluate whether explanations provided by \hype can help users more effectively identify relevant documents in realistic search scenarios, we conduct a human reranking experiment via Amazon Mechanical Turk (AMT). 
We prepare two conditions for comparison: (1) a title-only setting and (2) a title+path setting, where the title is shown along with a hierarchical category path explanation generated by \hype. 
For each query, five candidate documents are shown in both conditions, with the same title across settings; only the presence or absence of the category path differs, allowing for a controlled comparison of explanation impact.
We randomly sample 100 query-document instances from the NQ320K dev set where the title docid baseline with \hype successfully retrieves the gold document within the top-5 results.
Human judges are asked to (1) rank the five candidates based on their relevance to the query (i.e., human reranking), and (2) indicate their confidence in the ranked list using a 5-point Likert scale.
Based on the collected responses, we compute three metrics: Recall@1, which indicates whether the gold document was ranked first; MRR@5, which reflects how highly the gold document was ranked; and \textit{Confidence}, which measures how certain participants are in their rankings. 
This setup allows us to quantitatively assess whether the explanations produced by \hype improve both the accuracy and certainty of user decisions in realistic, information-limited search environments.
We show the interface for the human reranking in Figure~\ref{fig:humanrerankinginterface}

% With human-reranking results, we measure performance with Recall@1, MRR@5 and \textit{Confidence}.

% \subsection{Prompts}
% We intentionally did not consider semantic docids(+ HyPE) in our experiments. This is because semantic docids are constructed based on techniques such as hierarchical clustering, and thus inherently embed a semantic structure. Given that these structures are already formed in a coarse-to-fine manner, prepending hierarchical category paths to them would contradict the coarse-to-fine principle we defined in lines 148 and 158.

% \input{074humanevalinterface}

\subsection{Implementation Details}
\label{appendix:Implementationdetail}
We use T5-base~\cite{JMLR/Raffel2020/T5} as our backbone model. 
For the input of the indexing task, we utilize the FirstP approach as our document representations (Section~\ref{sec:preliminary}).
Additionally, for the indexing task, we employ five synthetic queries, generated by  using docT5query~\cite{NogueiraL19/docT5query} with nucleus sampling with parameters $p=0.8$ and $t=0.8$. 
We use new [DOC] token to separate the path from the docid, which we insert between the path and the docid. 
We optimize our model as described in \ref{subsubsec:modeloptimization}, while employing AdamW optimizer with a learning rate of 5e-4 and a batch size of 128, for up to 1M training steps. 
During the inference of \hype, we adopt \textit{\pars} strategy;
for the path generation stage, we generate three category paths (i.e., $K_p=3$), and for the docid generation stage, we use constrained beam search with a beam size of 100 (i.e., $m=100$). 
To build the summary docid baseline and keyword docid baseline, we utilize the off-the-shelf text summarization model based on BART~\cite{ACL/Liu2020/BART} and the keyword extraction tool~\cite{grootendorst2020keybert}.

\subsection{Ablation Study on the Number of Paths}
\label{appendix:ablationforpaths}
\begin{table}[t]
\centering
\small
\resizebox{.99\columnwidth}{!}{
\begin{tabular}{lccc}
\toprule
\textbf{Method} & R@10 & R@100 & M@100 \\
\midrule

Title docid & 78.7 & 89.3 & 68.6 \\
+ \hype($K=1$) & 82.4 & 89.5 & 70.0 \\
+ \hype($K=3$) & 83.5 & 90.1 & 71.0 \\
\midrule

Keyword docid & 77.1 & 85.5 & 67.6 \\
+ \hype($K=1$) & 78.5 & 85.1 & 67.1 \\
+ \hype($K=3$) & 79.1 & 86.2 & 67.7 \\
\midrule

Summary docid & 78.8 & 84.1 & 67.7 \\
+ \hype($K=1$) & 79.2 & 84.2 & 67.9 \\
+ \hype($K=3$) & 79.6 & 85.2 & 68.3 \\
\midrule

Atomic docid & 83.5 & 89.3 & 72.2 \\
+ \hype($K=1$) & 84.0 & 89.6 & 70.7 \\
+ \hype($K=3$) & 84.2 & 90.2 & 71.9 \\

\bottomrule
\end{tabular}
}
\caption{Retrieval accuracy across different docid types with varying numbers of paths on NQ320K.}
\label{tab:path_analysis}
\end{table}

We conduct additional experiments to examine how retrieval performance varies with the number of category paths considered.
As shown in Table~\ref{tab:path_analysis}, we observe consistent gains over the docid-only setting not only when using multiple paths but also in the single-path setting ($K=1$).
Moreover, incorporating additional paths further boosts performance across all baselines, revealing a clear gap between the without-\pars ($K=1$) and with \pars ($K=3$). 
% This provides empirical evidence that the observed improvements are indeed attributable to the hierarchical category paths themselves and not merely to an increased number of candidates.

\subsection{Analysis of Efficiency}
\label{appendix:efficientanalysis}
To quantify the inference cost introduced by generating hierarchical category paths, we measure the average inference time per instance using an NVIDIA RTX 4090 GPU. 
Specifically, we compare two decoding settings: (1) decoding only the docid, and (2) decoding both the docid and a single hierarchical category path. 
Our results show that the additional decoding required for generating a single path introduces only a marginal increase in inference time, demonstrating that \hype’s explainability can be achieved with minimal efficiency loss.

\subsection{Prompt}
Table~\ref{tab:reasoning_prompt} shows the prompt used to construct the path candidate set for the document with LLM.

\begin{table*}[ht]
    \centering
    \begin{tabular}{p{0.9\linewidth}} % Use p{} for paragraph formatting
    \toprule
    \textbf{Prompt: Select candidate path set for document} \\
    \midrule
    You're a taxonomy expert. 
    You will receive a document along with a set of candidate taxonomy hierarchy paths for the document. 
    Your task is to select the path that can represent the document. Exclude paths that are too broad or less relevant or contain too specific information such as year. \\
    You may list up to 3 paths, using only the paths in the candidate set. Do not include any explanation. \\ \\

    <Document title>:
    \{Document title\} \\

    <Document contents>: 
    \{Document contents\} \\

    <Candidate hierarchy paths>: 
    \{pre-candidate path set\} \\

    <Selected hierarchy paths>: 
    \{Candidate path set\} \\ 
    \bottomrule
    \end{tabular}
    \caption{The prompt for building final candidate path set.}
    \label{tab:reasoning_prompt}
\end{table*}

\end{document}